\documentclass[reprint,superscriptaddress,amsmath,amssymb,aps,prb,longbibliography]{revtex4-2}
\usepackage[dvipsnames]{xcolor}
\definecolor{mblue}{RGB}{31, 119, 180}
\usepackage{hyperref} 
\hypersetup{backref,
colorlinks=true,breaklinks,urlcolor=mblue,linkcolor=mblue,citecolor=mblue}
\usepackage{physics, braket, bm, amsthm, amsmath, amssymb, mathrsfs, graphicx, dcolumn, multirow}
\usepackage[]{ulem}

\renewcommand{\arraystretch}{1.2}
\usepackage{xcolor}

\begin{document}

\title{ Tighter thermalization bounds for perturbed quantum many-body scars}
\author{Meng-Yun Mao}
\affiliation{College of Physics, Nanjing University of Aeronautics and Astronautics, Nanjing 211106, China}
\affiliation{Key Laboratory of Aerospace Information Materials and Physics (NUAA), MIIT, Nanjing 211106, China}

\author{Zhixiang Sun}
\email{zsun@tju.edu.cn}
\affiliation{Center for Joint Quantum Studies and Department of Physics, School of Science, Tianjin University, 300072 Tianjin, China}

\author{Wen-Long You}
\email{wlyou@nuaa.edu.cn}
\affiliation{College of Physics, Nanjing University of Aeronautics and Astronautics, Nanjing 211106, China}
\affiliation{Key Laboratory of Aerospace Information Materials and Physics (NUAA), MIIT, Nanjing 211106, China}

\date{\today}

\begin{abstract}
Quantum many-body scars (QMBS) are exceptional eigenstates that defy thermalization, enabling long-lived coherent dynamics in strongly interacting systems. However, their stability under perturbations remains inadequately understood. In this work, we derive improved lower bounds on the thermalization time of QMBS under local perturbations with strength $\lambda$. Using both numerical simulations and analytical reasoning, we show that exact QMBS exhibit slow thermalization, with a timescale scaling as $\tau \sim \mathcal{O}(\lambda^{-1/d})$ owing to the stabilizing restricted spectrum-generating algebra (RSGA), which is a significant improvement over previous bounds (e.g., $\tau \sim \mathcal{O}(\lambda^{-1/(d+1)})$).
Counterintuitively, approximate QMBS can thermalize even more slowly under generic perturbations, exhibiting $\tau \sim \mathcal{O}(\lambda^{-2})$ scaling due to second-order perturbative effects in the absence of such protective structure. 
These distinct thermalization behaviors clarify how exact and approximate scars maintain coherence. Our work advances previous findings by establishing a tighter bound on the thermalization time,
clarifying when scarred dynamics remain long-lived under weak but generic perturbations.
\end{abstract}

\maketitle

\section{Introduction}
Quantum many-body scars (QMBS) have recently attracted considerable  attention for their role in challenging conventional paradigms of thermalization in isolated quantum systems. First observed in Rydberg atom arrays~\cite{bernien_probing_2017}, 
QMBS manifest as long-lived coherent oscillations for certain initial states, which represent a form of weak ergodicity breaking. This contrasts with the rapid thermalization predicted by the eigenstate thermalization hypothesis (ETH), which posits that in generic non-integrable systems, the expectation values of local observables should match those in the corresponding Gibbs ensemble in the thermodynamic limit—independent of the initial conditions.
These atypical dynamics of QMBS are linked to a small subset of special, non-thermal eigenstates embedded within an otherwise thermalizing spectrum. 
The PXP model, describing the constrained dynamics in Rydberg atom arrays~\cite{serbyn_quantum_2021}, serves as a paradigmatic example of how QMBS emerge in specific quantum systems.  
Since this seminal discovery, research on QMBS has progressed significantly, extending across a variety of physical systems, including spin chains~\cite{PhysRevLett.123.147201, Zhang_2024}, constrained lattice models~\cite{PhysRevResearch.3.043156, PhysRevB.101.024306, PhysRevResearch.4.013103, PhysRevB.108.195133}, and various experimental platforms—ranging from  ultracold atomic arrays~\cite{PhysRevResearch.5.023010, bluvstein_controlling_2021} and superconducting qubits~\cite{zhang_many-body_2023} to optical lattices~\cite{PhysRevLett.124.160604}. 

Despite being observed in a wide variety of physical systems, the stability of scarring under perturbations to the Hamiltonian and the underlying mathematical structures vary significantly across different models.
Numerous systems feature finely tuned algebraic structures that promote scarring, and in many instances, exact QMBS are associated with well-defined frameworks, such as restricted spectrum-generating algebras (RSGA)~\cite{PhysRevB.101.195131, PhysRevB.102.085140}. This approach seeks to identify an operator $\hat{Q}^\dagger$ that forms a simple commutation relation with the Hamiltonian $\hat{H}_0$ within the scarred subspace, $[\hat{H}_0,\hat{Q}^\dagger] |\phi_0\rangle= \mathcal{E} \hat{Q}^\dagger  |\phi_0\rangle$.  The repeated action of $\hat{Q}^\dagger$ on an eigenstate generates a tower of scarred states with 
equidistant energy spacing $\mathcal{E}$, leading to periodic quantum revivals in the system's dynamics.  For example, the AKLT model~\cite{PhysRevB.98.235155} and the spin-1 XY model~\cite{PhysRevLett.123.147201} feature exact QMBS that can be explained by a hidden $\mathfrak{su}(2)$ algebraic structure~\cite{PhysRevB.101.024306,hu2025krylovcomplexityquantummanybody}. The PXP model, however, presents a more intricate case. Although some of its eigenstates have exact matrix product state (MPS) representations~\cite{PhysRevLett.122.173401}, the broader scarring phenomenon is approximate. These approximate scars exhibit imperfect, long-lived revivals, which have been modeled using a  broken 
$q$-deformed $\mathfrak{su}(2)$ algebra~\cite{PhysRevB.106.205150}. These imperfect revivals can be stabilized by introducing auxiliary terms to the Hamiltonian~\cite{PhysRevLett.122.220603}, revealing an emergent $\mathfrak{su}(2)$ symmetry that protects the scar states. 

While the fundamental characteristics of QMBS are increasingly understood, the stability of these scars under perturbations that disrupt these structures remains poorly understood, and continues to be a crucial topic of ongoing research~\cite{PhysRevB.103.104302, PhysRevB.98.155134, PRXQuantum.2.030349, PhysRevB.104.214305, Lin2020, PhysRevLett.125.230602, Halimeh2023robustquantummany, chioquetta_stability_2025, PhysRevX.15.011020}. 
Despite being a vanishing fraction of states in a Hilbert space dominated by thermal states, QMBS are expected to exhibit quasiperiodic revivals under weak perturbations. However, sufficiently strong perturbations can disrupt this behavior and induce thermalization~\cite{PRXQuantum.4.040348, PRXQuantum.2.030349, PhysRevB.104.214305}.  Recent studies further suggest that QMBS are stable under first-order perturbations but can become unstable due to higher-order perturbative corrections arising from the hybridization of exact scar eigenstates with thermal eigenstates~\cite{PRXQuantum.4.040348}.
While qualitative insights abound, comprehensive quantitative investigations into the thermalization time $t^{*}$ of QMBS remain scarce. For instance, a foundational work established bounds for exact QMBS under perturbation~\cite{Lin2020}, providing a specific form of  $t^{*} \sim \mathcal{O}(\lambda^{-1/(d+1)})$, where $\lambda$ is the perturbation strength and $d$ is the dimension of the system.   This bound is inherently connected to how information spreads, as constrained by locality and the Lieb-Robinson bound~\cite{lieb_finite_1972, bru2016lieb}. Other studies have provided specific thermalization time bounds for particular models quenched from special initial states, such as   $t^{*} \sim \mathcal{O}(\lambda^{-2})$ for both the spin-1/2 XYZ model~\cite{PhysRevB.105.L060301} and the spin-1 Kitaev model~\cite{PhysRevB.108.104411}.    
Despite these valuable efforts, a unified and general conclusion regarding thermalization time for QMBS remains elusive, highlighting a significant gap in current understanding. Our study addresses this gap by investigating the stability of both exact and approximate QMBS under various perturbations.

In this work,  we derive a tighter bound on the thermalization time of systems exhibiting QMBS under weak perturbations, compared to the bound presented in Ref.~\cite{Lin2020}. Importantly, our result applies not only when the initial state is a scarred eigenstate but also extends to certain specially prepared initial states. The remainder of the paper is organized as follows.  In Sec.~\ref{sec2},  we derive a universal lower bound on the thermalization time, showing that it scales as $t^{*} \sim \mathcal{O}(\lambda^{-1/d})$ for systems with exact QMBS. Section~\ref{sec3} presents numerical examples that support our theoretical derivation. Finally, we conclude with a summary and discussion in Sec.\ref{sec4}.

 \section{Derivation of the Thermalization Time Bound}\label{sec2}

In this section, we analyze the dynamics of local observables in a $d$-dimensional system that hosts exact QMBS stabilized by the RSGA, subject to weak local perturbations. From this setup we derive a universal lower bound on the thermalization time. 

Consider a chain with a local, finite-range Hamiltonian $\hat H_0$ that supports a scarred subspace. We perturb it by a weak local operator
$\hat V=\sum_j \hat v_j$, where each $\hat v_j$ is supported in a neighborhood of site $j$ and has bounded norm (we absorb overall scales into the coupling parameter $\lambda$). The full Hamiltonian is
\begin{eqnarray}
\hat H=\hat H_0+\lambda \hat V,\qquad 0<\lambda\ll 1,
\end{eqnarray}
with $\lambda$ a dimensionless perturbation strength.
We analyze the system's evolution from an initial state $|\psi(0)\rangle$ that is a superposition of scar eigenstates of $\hat{H}_0$, which are known to have equal energy spacing. 

The nonthermal dynamics are captured by the deviation $\delta m_0(t)$ between the expectation value of the local observable $\hat{m}_{0}$ under the perturbed evolution and that under the unperturbed evolution at time $t$, which is defined as
\begin{eqnarray}
    \delta m_0(t) & = & \langle \psi(0) | e^{i \hat{H} t} \hat{m}_{0} e^{-i \hat{H} t} | \psi(0) \rangle \nonumber \\ 
    & - & \langle \psi(0) | e^{i \hat{H}_{0} t} \hat{m}_{0} e^{-i \hat{H}_{0} t} | \psi(0) \rangle.
\end{eqnarray}
We have set the reduced Planck constant to $\hbar = 1$. With aid of the Lieb-Robinson bound, the deviation $\delta m_0(t)$ can be bounded as
\begin{equation}
  \label{eq:deltam}
    | \delta m_0(t) | \leq \lambda (c_0 + c_1 t^d),
\end{equation}
where $c_0$ and $c_1$ are $\mathcal{O}(1)$ constants independent of $t$ and $\lambda$, as detailed in the Supplemental Material (SM)~\cite{SM} (see also references ~\cite{Lin2020, PhysRevB.101.195131, PhysRevB.102.085140, cheneau_light-cone-like_2012, PhysRevLett.117.091602, maldacena_bound_2016, hashimoto_out--time-order_2017, rigol_thermalization_2008, mallayya_prethermalization_2019, PhysRevB.104.184302} therein). For the expectation value of the observable at moment $t$ to change by an $\mathcal{O}(1)$ amount, signifying thermalization, the right-hand side must be of $\mathcal{O}(1)$. 
The term $\lambda c_0$ remains small for weak perturbations ($\lambda \ll 1$) and does not grow in time, so it cannot by itself signal thermalization. 
Thus, the dominant term for large $t$ is $c_1 \lambda t^d$. This implies that the system will not thermalize before a timescale $t^{*}$ given by:
\begin{equation} \label{eq:t_star_scaling}
t^{*} \sim \mathcal{O}(\lambda^{-1/d}).
\end{equation}
This establishes a parametrically long prethermal regime where scarred dynamics persist.
For systems in one dimension, the thermalization time bound scales as $\mathcal{O}(\lambda^{-1})$.
 
\begin{table*}[t] 
  \renewcommand{\arraystretch}{1.3} %
  \caption{\textbf{Summary of QMBS models studied in this work.} 
  We list boundary conditions, observables, perturbations, QMBS type, and the resulting thermalization times.
  We argue that the thermalization time scales as $\mathcal{O}(\lambda^{-1})$ for exact QMBS and $\mathcal{O}(\lambda^{-2})$ for approximate QMBS, 
  where $\lambda$ denotes the perturbation strength.}  
  \label{table_summary}
  \centering
  \begin{tabular}{@{\hskip 4pt}c@{\hskip 4pt} l@{\hskip 6pt} c@{\hskip 6pt} l@{\hskip 6pt} c@{\hskip 4pt} c@{\hskip 6pt}}  
      \hline\hline
     \textbf{Model} & \textbf{Boundary}  & \textbf{Observable} $m_0$ &  \textbf{Perturbation} & \textbf{QMBS Type} & \textbf{Thermalization Time}\\
      \hline
   Spin-1 XY model (\ref{xy}) & PBC  & $\hat{O}^{+-}$ [Eq.(\ref{operatorO+-})] & $ \lambda \sum_jS_{j}^{z} S_{j+1}^{z}$ & Exact & $\mathcal{O}(\lambda^{-1})$ \\
    Spin-1 XY model (\ref{xy}) & OBC & $\hat{O}^{+-}$ [Eq.(\ref{operatorO+-})] & $ \lambda \sum_jS_{j}^{z} S_{j+1}^{z}$  & Exact & $\mathcal{O}(\lambda^{-1})$  \\
    Deformed PXP model (\ref{Brus}) & PBC & $\hat{O}^{z} $ [see Eq.~(\ref{operatorOzz})] & $ \lambda \sum_j\sigma_j^z$ & Exact & $\mathcal{O}(\lambda^{-1})$ \\
    Deformed PXP model (\ref{Brus})  & OBC & $\hat{O}^{z}$ [see Eq.~(\ref{operatorOzz})]  & $ \lambda \sum_j\sigma_j^z$  & Approximate & $\mathcal{O}(\lambda^{-2})$ \\
 Spin-1 Kitaev model (\ref{Kitaev}) & PBC & $\hat{O}^{+-}$ [Eq.(\ref{operatorO+-})] & $ \lambda \sum_j(S_{j}^{z})^2$  & Approximate & $\mathcal{O}(\lambda^{-2})$ \\
 DWC model (\ref{ham:DWC}) & PBC  & $\hat{O}^{x} $ [see Eq.~(\ref{operatorOzz})] & $ \lambda \sum_j\sigma_j^x$ & Exact & $\mathcal{O}(\lambda^{-1})$\\
Deformed DWC model (\ref{ham:dDWC}) & PBC & $\hat{O}^{x}$ [see Eq.~(\ref{operatorOzz})] & $ \lambda \sum_j\sigma_j^z$  & Approximate & $\mathcal{O}(\lambda^{-2})$ \\
      \hline\hline
  \end{tabular}
\end{table*}

\section{Numerical Validation of the Thermalization Bound} \label{sec3}
In this section, we provide numerical validation for the analytical thermalization bounds derived in Sec.~\ref{sec2}, specifically focusing on one-dimensional (1D) systems. We investigate two distinct classes of systems to test the scaling of the thermalization time $\tau$ with the perturbation strength $\lambda$. 
Our central finding reveals a clear dichotomy: models hosting \textit{exact} QMBS consistently exhibit a thermalization timescale $\tau \sim \mathcal{O}(\lambda^{-1})$, while those with \textit{approximate} QMBS thermalize  more slowly, with quadratic scaling  $\tau \sim \mathcal{O}(\lambda^{-2})$. These scalings are consistent with the rigorous lower bound $t^{*} \sim \mathcal{O}(\lambda^{-1})$ proved for exact scars.
 
We simulate the quantum quench dynamics for each model using exact diagonalization (ED). The time evolution is initiated from states known to exhibit prominent revivals characteristic of QMBS. Importantly, across all models considered, the canonical scar initial states we employ are fixed by the underlying algebraic/constrained structure and are independent of microscopic Hamiltonian parameters. To quantify the thermalization timescale $\tau$, we fit the revival envelope of local observables. We consider two fitting Ans\"atze for the decay: an exponential form, $A e^{-t/\tau}\cos(\omega t)$, and a Gaussian form, $A e^{-(t/\tau)^2}\cos(\omega t)$. For each model, we select the form that best describes the data.
 
Throughout, we distinguish two timescales. The rigorous thermalization bound $t^*$ denotes the earliest time at which an $\mathcal{O}(1)$ deviation of a local observable can occur, i.e., a prethermal lifetime lower bound. The fitted parameter $\tau$ is the operational decay time extracted numerically from revival envelopes. While $\tau$ is not identical to $t^*$, their scalings with~$\lambda$ are consistent across all models studied. 
For completeness, we present results under both periodic (PBC) and open (OBC) boundary conditions. A summary of the main findings is provided in Tab.~\ref{table_summary}.

{\it The spin-1 XY model.--}We begin our numerical validation with the spin-1 XY model, a system known to host exact QMBS~\cite{PhysRevLett.123.147201}. The Hamiltonian on an $L$-site chain is given by:
\begin{eqnarray}\label{xy}
  \hat{H}_{\rm XY} & = & \sum_{i=1}^{L} \Big[J (S_i^xS_{i+1}^x+S_i^yS_{i+1}^y) + h S_i^z + D(S_i^z)^2 \nonumber \\
  & & + J_3 (S_i^xS_{i+3}^x+S_i^yS_{i+3}^y)  \Big], 
\end{eqnarray}
where the parameters $J$, $h$, $D$, and $J_3$ respectively denote the strength of the nearest-neighbor XY interactions, the longitudinal magnetic field, the single-ion anisotropy, and the third-nearest-neighbor XY coupling.  For our simulations, we set these parameters to $(J,h,D,J_3) = (1,1,0.1,0.1)$.

We initialize the system in the nematic N\'{e}el state. For the 1D chain, this state is defined as:
\begin{equation} \label{eq:initial_state_xy}
\ket{\psi(0)} = \bigotimes_{j=1}^{L} \frac{\ket{+1}_j - (-1)^j \ket{-1}_j}{\sqrt{2}},
\end{equation}
where $\ket{\pm 1}_j$ denotes the eigenstate of the local spin operator $S_j^z$ with eigenvalue $\pm 1$. This initial  state is specifically  chosen because it is a known superposition of the model's exact scar eigenstates and thus exhibits perfect revivals in the unperturbed dynamics. To study the stability of these revivals, we introduce a perturbation of the form $\lambda \hat{V} = \lambda \sum_{i}S_i^zS_{i+1}^z$, governed by the strength $\lambda$.

\begin{figure}[!t]
  \begin{center}
  \includegraphics[width=\columnwidth]{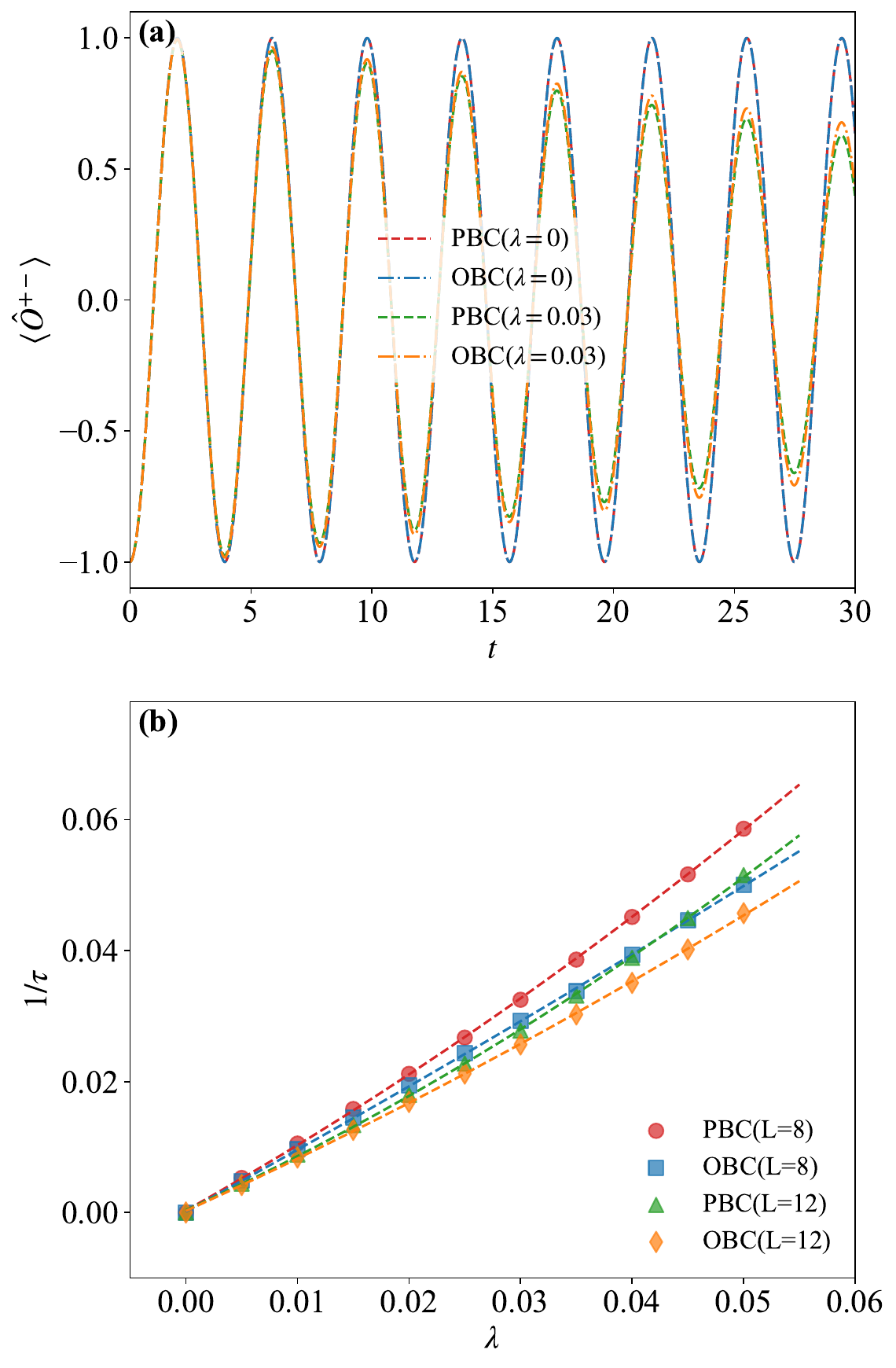}
  \end{center}
   \caption{\label{fig_xy} 
   (a) Dynamics of the local observable $\langle \hat{O}^{+-}\rangle$ following a quench from the nematic N\'eel state in the spin-1 XY model with $L=12$. The dynamics are shown for the unperturbed case ($\lambda=0$) and a perturbed case ($\lambda=0.03$) under both  PBC and OBC.
   (b) Extracted thermalization time $\tau$ as a function of perturbation strength $\lambda$ for system sizes $L=8, 12$ under different boundary conditions. Symbols represent numerical data obtained from Gaussian fits, while the dashed lines show the second-order polynomial fits of the decay rate $1/\tau$. In the weak-perturbation regime,  the leading dependence is approximately linear, $1/\tau \propto \lambda$.
   }
\end{figure}

The expectation value of the local observable 
\begin{equation} \label{operatorO+-}
\hat{O}^{+-}=\frac{1}{2} [(S_1^+)^2+(S_1^-)^2]
\end{equation}
is plotted as a function of time in Fig.~\ref{fig_xy} (a) for various boundary conditions and perturbation strengths, where $|\psi(t)\rangle$ represents the time-evolved state. In the unperturbed case ($\lambda=0$), $\langle \hat{O}^{+-} \rangle$ exhibits perfect revivals, a hallmark of the underlying scar states. As expected, introducing the perturbation suppresses these revivals and induces thermalization. To quantify this effect, the evolution of $\langle \hat{O}^{+-} \rangle$ under perturbation is fitted with a Gaussian decay function of the form $A e^{-t^2/\tau^2} \cos(\omega t)$. 
The extracted thermalization time, $\tau$, is plotted as a function of the perturbation strength $\lambda$ in Fig.~\ref{fig_xy}(b). 
In the weak-perturbation regime, the decay rate $1/\tau$ is approximately linear in $\lambda$. Note that the numerical data exhibit a slight curvature, which we attribute to subleading corrections beyond the leading linear behavior; a more detailed fitting analysis is provided in SM~\cite{SM}. 
This implies a characteristic timescale that scales as $\tau \sim \mathcal{O}(\lambda^{-1})$. We note that this scaling differs from the $\mathcal{O}(\lambda^{-2})$ dependence argued for in Ref.~\cite{Lin2020}, which was based on an exponential decay fit. However, our proposed $\mathcal{O}(\lambda^{-1})$ scaling appears to be more consistent not only with our findings but also with the numerical data  for the weak perturbation regime.

This linear scaling of the decay rate with perturbation strength is a robust feature, holding for both PBC and OBC. The linear dependence traces back to the structure of the scar eigenstates at $\lambda=0$, where boundary terms modify the energy eigenvalues, while the eigenstates are determined solely by the magnon number and remain unchanged in both cases. Consequently, because the nematic N\'eel state (or indeed, any superposition of the scar states) is an exact eigenstate superposition in both cases, the perfect revivals at $\lambda=0$ are independent of the boundary conditions. While the qualitative scarring behavior is therefore identical, a quantitative difference in stability emerges under perturbation. As shown in Fig.~\ref{fig_xy}(b), the slope of the $1/\tau$ vs.~$\lambda$ curve is consistently smaller for OBC.
This indicates a greater resilience of the scar dynamics in the open-chain system. We attribute this to the additional interaction pathway between the first and last spins that is present under PBC. This extra link provides another channel for perturbations to propagate through the system, thereby accelerating decoherence and leading to a faster decay of the revivals.

{\it The deformed PXP model.--}As a second example of exact QMBS, we examine the deformed PXP model. This model is particularly instructive as it is built upon the standard PXP model---a paradigmatic system for approximate scars---by adding specific long-range interactions that stabilize the scarring phenomenon and render it exact~\cite{PhysRevLett.122.220603}.

The standard PXP model Hamiltonian is given by:
\begin{equation}\label{pxp_ham}
  \hat{H}_{\rm PXP} = \sum_{i=1}^L P_{i-1}\sigma_i^xP_{i+1},
\end{equation}
where $\sigma_i^x$ is the Pauli-X operator at site $i$, and $P_i = (1-\sigma_i^z)/2$ is the projector onto the local down-spin state ($\ket{\downarrow}$). This projector enforces the Rydberg blockade constraint, which forbids adjacent up-spins ($\ket{\uparrow}$).

To convert the approximate scars of this model into exact ones, a deformation term, $\hat{H}_\delta$, is introduced. This term adds a set of finely-tuned, long-range interactions given by:
\begin{equation}\label{Hdeltaham}
 \hat{H}_\delta = - \sum_{i=1}^{L} \sum_{\delta=2}^{R} h_\delta P_{i-1} \sigma^{x}_{i} P_{i+1} (\sigma^{z}_{i-\delta} + \sigma^{z}_{i+\delta}).
\end{equation}
Here, the interaction strength $h_\delta$ depends on the distance $\delta$ according to the relation $h_\delta = h_0 (\phi^{\delta-1} + \phi^{\delta+1})^{-2}$, with $h_0 \approx 0.051$ and $\phi$ being the golden ratio. Following Ref.~\cite{PhysRevLett.122.220603}, we truncate the interaction range at $R = L/2$. The full Hamiltonian for the deformed PXP model is then the sum of these two components:
\begin{equation}\label{Brus}
  \hat{H}_{\rm dPXP} = \hat{H}_{\rm PXP} + \hat{H}_\delta.
\end{equation}
The crucial effect of the $\hat{H}_\delta$ term is to transform the imperfect, quasiperiodic revivals of the standard PXP model into the perfect, stable revivals characteristic of exact QMBS~\cite{PhysRevLett.122.220603}. 
 
\begin{figure}[!t]
\begin{center}
\includegraphics[width=\columnwidth]{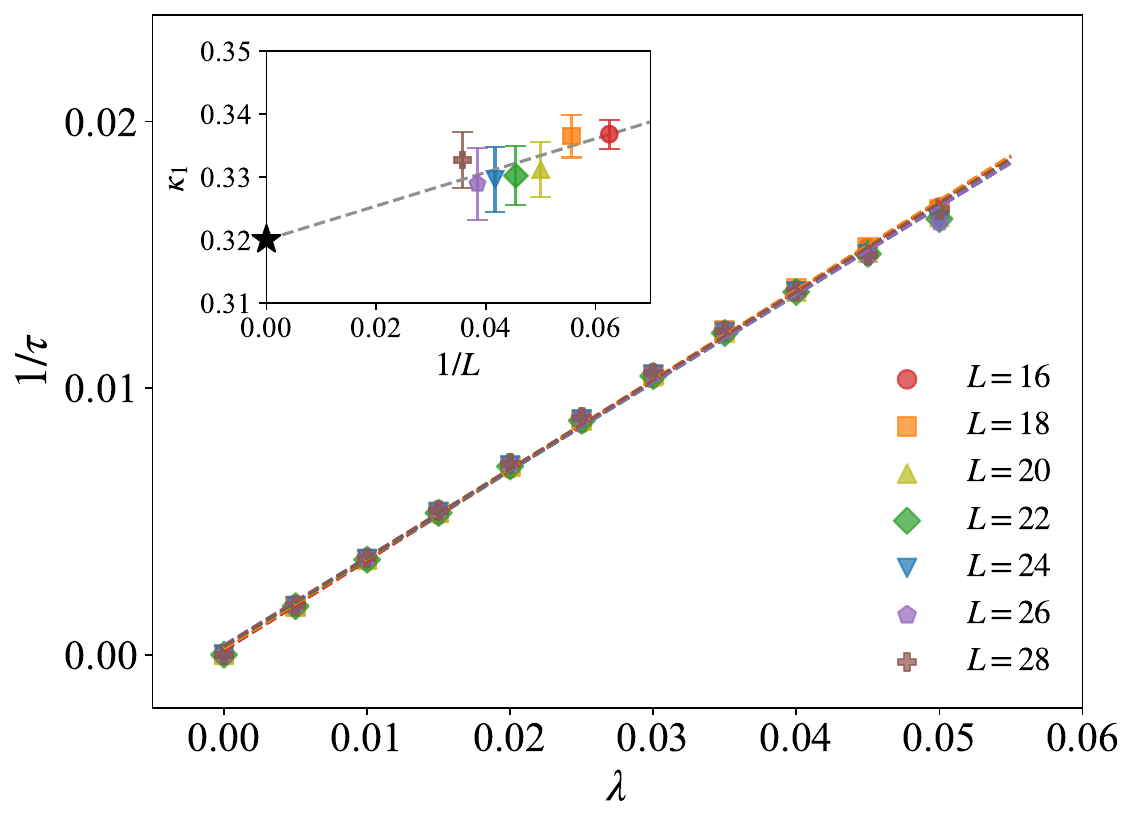}
\end{center}
\caption{\label{fig_pxp_perfect} 
Decay rate $1/\tau$ for the local observable $\langle \hat{O}^{z} \rangle$ in the deformed PXP model under PBC, plotted as a function of perturbation strength $\lambda$. 
The data for different system sizes are fitted to the leading-order term $1/\tau = \kappa_1 \lambda$ (dashed lines), consistent with the linear scaling characteristic of exact QMBS.
Inset shows the scaling of the fitted slope $\kappa_1$ with inverse system size $1/L$. Error bars represent the uncertainty of the linear fits. The dashed line is a linear extrapolation to the thermodynamic limit ($1/L \to 0$), where the limiting value is indicated by the black star.}
\end{figure}

We then analyze the stability of these exact scars by introducing a local perturbation $\lambda \hat{V} = \lambda \sum_i \sigma^z_i$.  We consider the local staggered observable
\begin{equation} \label{operatorOzz}
\hat{O}^{\alpha}=\frac{1}{2} (\sigma^\alpha_1-\sigma^\alpha_2), \qquad \alpha\in\{x,z\}.
\end{equation}
The dynamics are initiated from the N\'{e}el state $|\psi(0)\rangle = | \downarrow \uparrow \downarrow \uparrow \dots \rangle$.   As shown in Fig.~\ref{fig_pxp_perfect}, we again find that the decay rate of the observable 
$\langle \hat{O}^{z} \rangle$ follows a clear linear dependence on the perturbation strength, $1/\tau = \kappa_1 \lambda$, consistent with our theoretical bound $t^{*} \sim \mathcal{O}(\lambda^{-1})$. The data for different system sizes nearly collapse onto a single line, indicating that this scaling behavior is robust and persists in the thermodynamic limit.

\begin{figure}[!t]
  \begin{center}
  \includegraphics[width=\columnwidth]{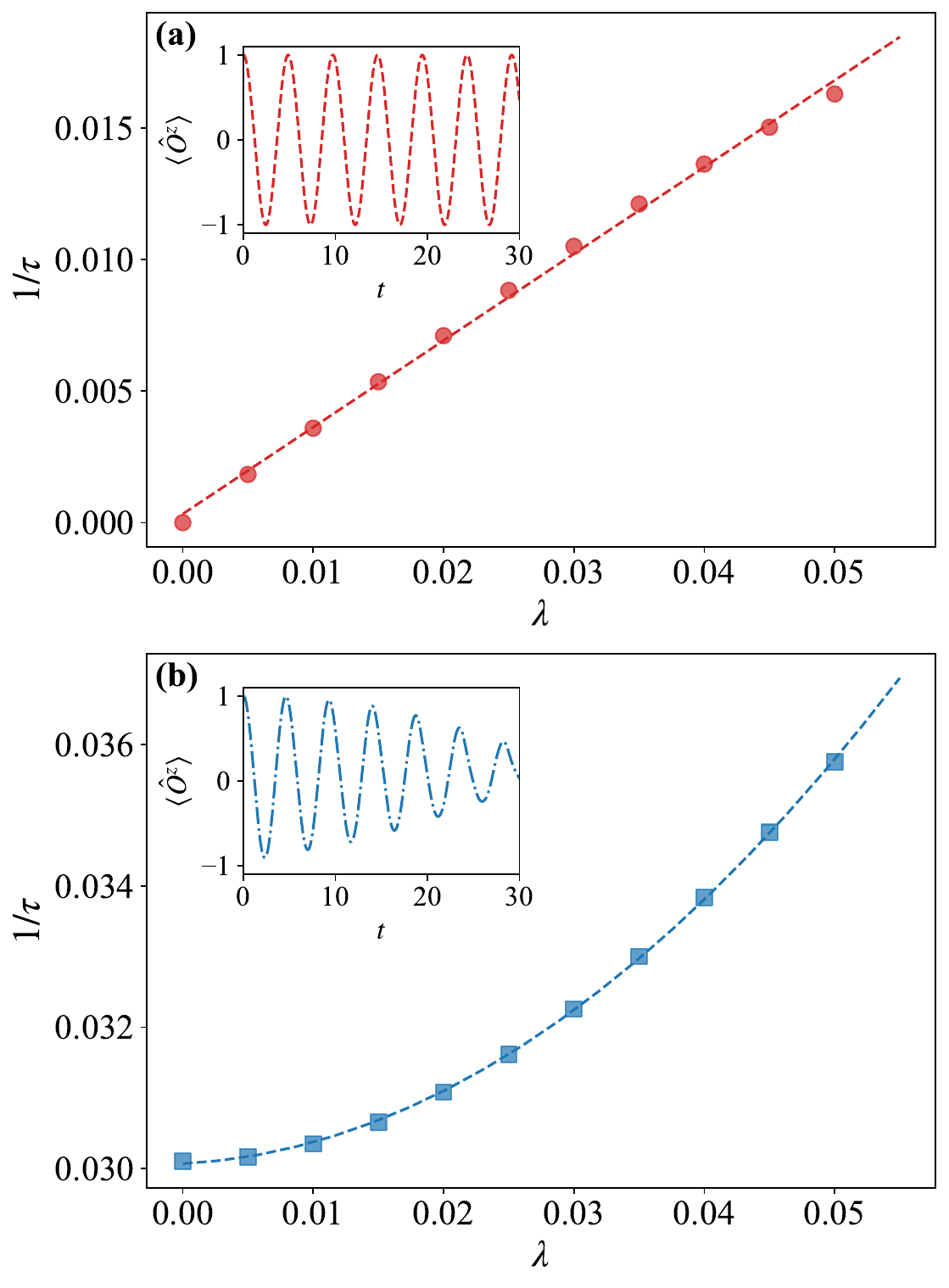}
  \end{center}
   \caption{\label{fig_pxp_perfect_boundary} 
   Comparison of thermalization dynamics in the deformed PXP model for $L=24$.
(a) Under PBC, the decay rate $1/\tau$ scales predominantly linearly with the perturbation strength $\lambda$. 
(b) Under OBC, the decay rate scales quadratically with the perturbation, as shown by the linear fit against $\lambda^2$.
Insets show the dynamics at $\lambda=0$, confirming the transition from exact (PBC) to approximate (OBC) scarring. 
  }
\end{figure}

Interestingly, the nature of the scarring in this model is highly sensitive to the boundary conditions. While the model hosts exact QMBS under PBC, switching to OBC (explicit Hamiltonian in SM~\cite{SM}) breaks the specific symmetries that protect the perfect revivals. As shown in the inset of Fig.~\ref{fig_pxp_perfect_boundary}, the dynamics under OBC at $\lambda=0$ are no longer perfectly periodic, signifying that the QMBS have become approximate. Consequently, the thermalization dynamics change dramatically. 
Figure~\ref{fig_pxp_perfect_boundary} shows that while the PBC case exhibits the expected $1/\tau \propto \lambda$ scaling at leading order, the OBC case is dominated by a quadratic trend, $1/\tau \propto \lambda^2$. 
This model therefore provides a controlled setting to observe the transition from exact to approximate scarring and its direct impact on thermalization, a conclusion we will reinforce in the next example. Furthermore, we extend our analysis to investigate the stability against different types of $k$-local perturbations in SM~\cite{SM}.

{\it The spin-1 Kitaev model.--}
Having introduced the PXP Hamiltonian, we now analyze the spin-1 Kitaev model, which hosts intrinsically approximate QMBS.  The Hamiltonian is given by:
\begin{equation}\label{Kitaev}
  \hat{H}_{\rm K} = \sum_{j=1}^{L/2} (S_{2j-1}^x S_{2j}^x + S_{2j}^y S_{2j+1}^y).
\end{equation} 
A key feature of this model is the existence of local conserved quantities, the bond-parity operators $\hat{W}_j$, defined as $\hat{W}_{2j-1} = \Sigma_{2j-1}^{y} \Sigma_{2j}^{y}$ and $\hat{W}_{2j} = \Sigma_{2j}^{x} \Sigma_{2j+1}^{x}$ for odd and even bonds, respectively. Here, the site parity operators are given by $\Sigma_i^{\alpha} = e^{i \pi S_i^{\alpha}}$ with $\alpha \in \{x,y,z\}$. The eigenvalues $w_j = \pm 1$ of these operators $\hat{W}_{j}$ act as $\mathbb{Z}_2$ gauge charges, fragmenting the entire Hilbert space into $2^L$ dynamically disconnected "flux sectors," each labeled by a configuration of these charges.
The significance of this model lies in the physics of its flux-free sector (where all $w_j=+1$), which contains the ground state. As demonstrated in Refs.~\cite{PhysRevResearch.4.013103, PhysRevB.108.104411}, within this specific constrained subspace, the Kitaev Hamiltonian [Eq.~\eqref{Kitaev}] can be exactly mapped to the standard spin-1/2 PXP model, whose Hamiltonian is given by Eq.~\eqref{pxp_ham}.

To study the stability of its scars, we introduce a single-ion anisotropy perturbation, $\lambda \hat{V}=\lambda \sum_{i} (S_i^z)^2$. Crucially, this specific perturbation, when projected into the flux-free sector and translated into the PXP language, maps precisely to a chemical potential or detuning term. Therefore, the full perturbed Hamiltonian, $\hat{H}_{\rm K} + \lambda \hat{V}$, restricted to the flux-free subspace, is unitarily equivalent to the well-known detuned PXP model. It is established that such a detuning term breaks the special structure of the pure PXP model, causing its approximate scar subspace to leak into the surrounding thermal states. This rigorous mapping thus provides a solid physical foundation for studying the stability of approximate QMBS.

\begin{figure}[!t]
  \begin{center}
  \includegraphics[width=\columnwidth]{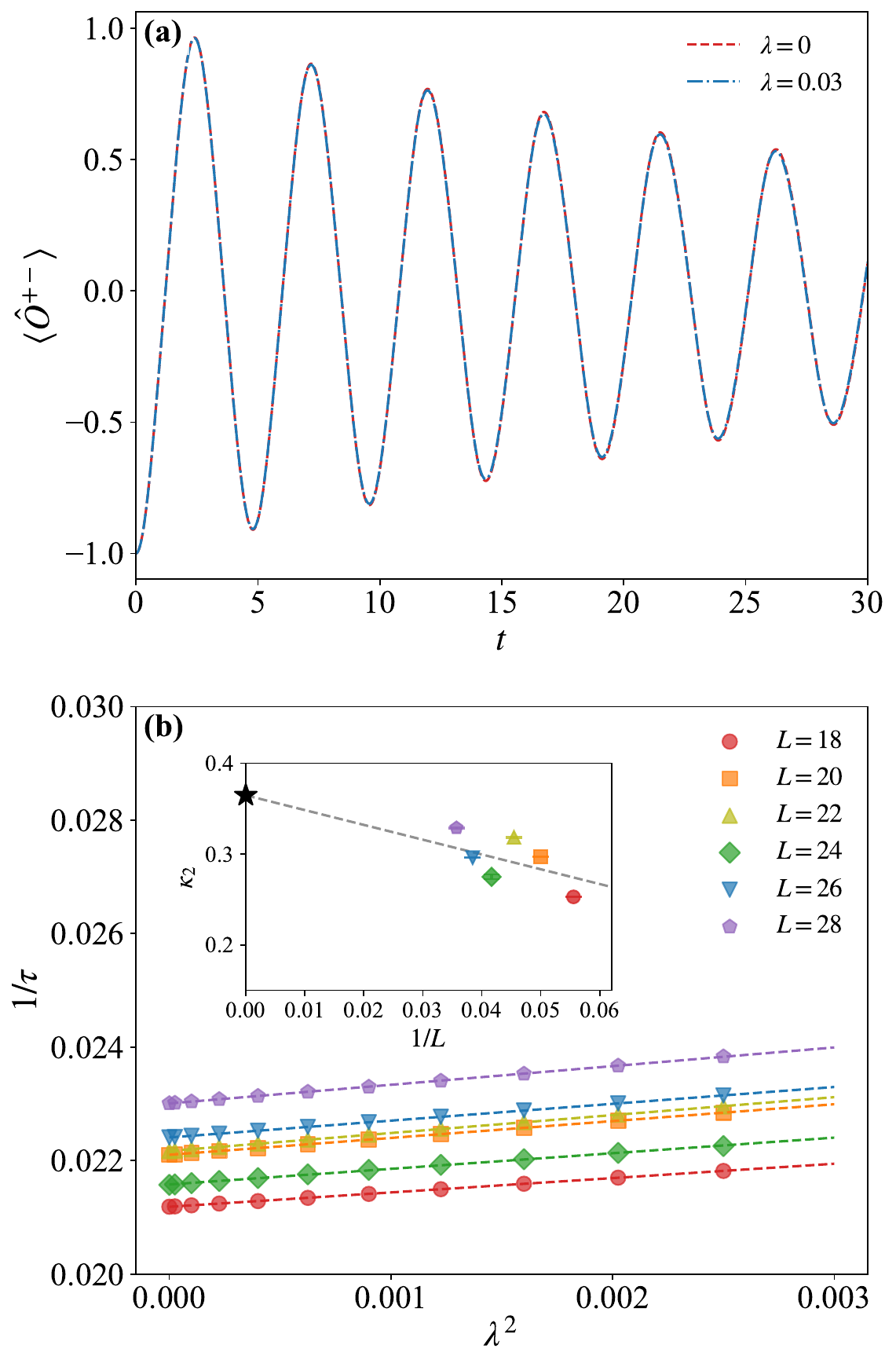}
  \end{center}
   \caption{\label{fig_Kitaev} 
   (a) Dynamics of the observable $\langle \hat{O}^{+-} \rangle$ in the spin-1 Kitaev model for $L=28$. The imperfect revival, even for the unperturbed case ($\lambda=0$), signifies approximate QMBS.
   (b) The decay rate $1/\tau$ plotted against $\lambda^2$ for different system sizes. The linear fit confirms the quadratic scaling $1/\tau =\kappa_2  \lambda^2$. 
   Inset shows the fitted slope $\kappa_2$ extrapolates to a non-zero value in the thermodynamic limit.
   }
\end{figure}

We initialize the dynamics in the state (\ref{eq:initial_state_xy}), which corresponds to the N\'eel state in the PXP language and resides within the flux-free sector. The resulting dynamics of the observable $\langle \hat{O}^{+-} \rangle$ are shown in Fig.~\ref{fig_Kitaev} (a). Even without the explicit perturbation ($\lambda=0$), the revivals are imperfect, which is an inherent feature of the mapping to the pure PXP model and confirms the approximate nature of the QMBS.

The extracted decay rate is plotted against $\lambda^2$ in Fig.~\ref{fig_Kitaev}(b). 
In stark contrast to the exact-scar cases, the data reveal a clear quadratic dependence, 
$1/\tau=\kappa_2\lambda^2$, implying a slower thermalization timescale 
$\tau\sim\mathcal{O}(\lambda^{-2})$. 
Unlike exact scars, where generic perturbations destroy the underlying algebraic structure and trigger a non-perturbative instability, approximate scars here are better understood as quasiparticle-like structures. Under weak perturbations, their relaxation is driven by incoherent scattering into the thermal continuum, leading to a decay rate $\lambda^2$ governed by Fermi's golden rule:
\begin{equation}
\frac{1}{\tau}\;\simeq\;2\pi\,\lambda^2\,\rho(E)\,\big|\langle f|\hat V|i\rangle\big|^2,
\end{equation}
where $\rho(E)$ is the many-body density of states at energy $E$. 
A detailed derivation of this result is provided in SM~\cite{SM}. 
The inset of Fig.~\ref{fig_Kitaev}(b) further shows that the fitted slope $\kappa_2$ 
converges to a finite value as $L\to\infty$, confirming that the observed scaling is robust 
in the thermodynamic limit.

{\it The DWC model and its deformation.--} 
To benchmark our analysis, we investigate the domain-wall-conserving (DWC), or xorX, model~\cite{feng_uncovering_2025}. This spin-1/2 system is notable for its exact scar states, which possess a known analytical form~\cite{PhysRevB.101.024306}. The Hamiltonian is given by:
\begin{equation} \label{ham:DWC}
  \hat{H}_{\textrm{DWC}} = \sum_{i=1}^{L} (\sigma_{i}^{x} - \sigma_{i-1}^{z} \sigma_{i}^{x} \sigma_{i+1}^{z}  + \Delta \sigma_{i}^{z} + J \sigma_{i}^{z} \sigma_{i+1}^{z}),
\end{equation}
where $\Delta$ and $J$  represent the strengths of the Zeeman field and Ising interaction terms, respectively. 
In the following numerical simulation, we set $\Delta = J = 1$.  Unlike the PXP model,  the DWC model enforces a local constraint on $X$ flips: a spin can only flip if its nearest neighbors are in opposite $\sigma^{z}$ states, resulting in the conservation of domain walls as $[\hat{H}_{\textrm{DWC}}, \sum_i \sigma_{i}^{z} \sigma_{i+1}^{z}]=0$.

This model hosts a tower of exact scar states $| \mathcal{S}_{n} \rangle$ that can be generated algebraically from the polarized vacuum $|\Omega\rangle=|\downarrow\downarrow\cdots\downarrow\rangle$ by a raising operator $\hat Q^\dagger$, 
given by $| \mathcal{S}_{n} \rangle$ = $\frac{1}{n!\sqrt{\mathcal{N}(L , n)}} (\hat{Q}^{\dagger})^{n} | \Omega \rangle$,
where $ | \Omega \rangle = | \downarrow \downarrow \dots \downarrow  \rangle$, $n = 0 , \dots , L/2$ is an integer,  $\mathcal{N}(L , n) = \frac{L}{n} \binom{L - n - 1}{n - 1}$, and the raising operator  $\hat{Q}^{\dagger}$ = $\sum_{j=1}^{L} (-1)^{j} P_{j-1} \sigma^{+}_{j} P_{j+1}$,
where $\sigma^{\pm}_{j} = (\sigma^{x}_{j} \pm i \sigma^{y}_{j}) / 2$. 
 We initialize the system in a specific superposition of these states, 
 \begin{eqnarray}
 \ket{\psi(0)} & = & \frac{1}{\sqrt{Z}} \prod_{j=1}^{L} [1 + (-1)^{j} P_{j-1} \sigma^{+}_{j} P_{j+1}] | \Omega \rangle \nonumber \\
  & = & \frac{1}{\sqrt{Z}} \sum_{n=0}^{L/2} \sqrt{\mathcal{N}(L ,n)} | \mathcal{S}_{n} \rangle,
\end{eqnarray}
where the normalization factor $Z = \frac{1}{\sqrt{5}}((\frac{1+\sqrt{5}}{2})^{L} - (\frac{1-\sqrt{5}}{2})^{L})$.
It exhibits strong revivals of the observable $\langle \hat{O}^{x}\rangle$. 

\begin{figure}[!t]
  \begin{center}
  \includegraphics[width=\columnwidth]{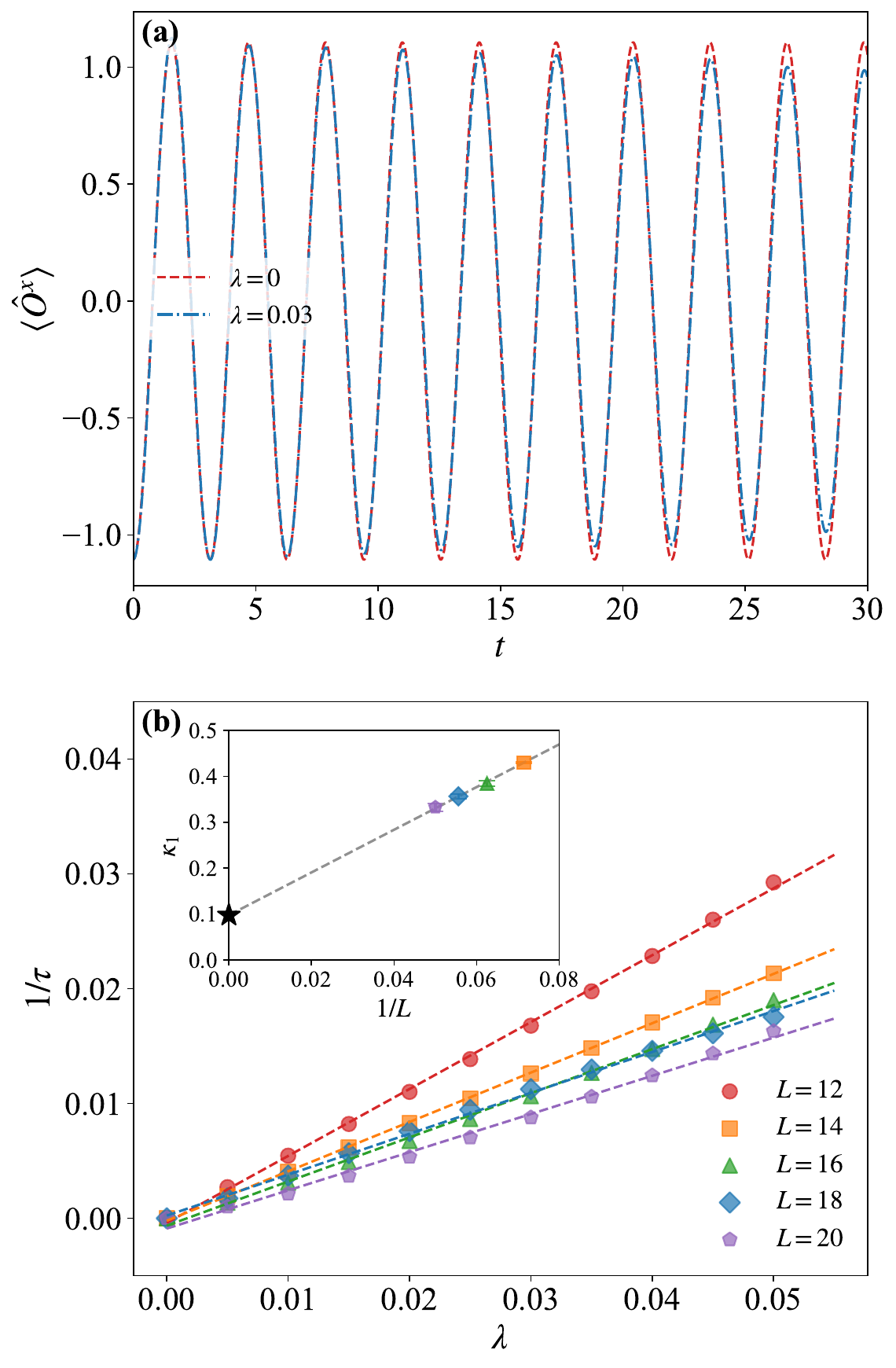}
  \end{center}
   \caption{\label{fig_xorX} 
   (a) Dynamics of $\langle \hat{O}^{x} \rangle$ in the exact DWC model for $L=20$, showing perfect revivals for the unperturbed case ($\lambda = 0$).
   (b) Decay rate $1/\tau$ versus perturbation strength $\lambda$ for various system sizes. The linear fits confirm the scaling $1/\tau = \kappa_1 \lambda$. Inset shows  the fitted slope $\kappa_1$ as a function of inverse system size.
   }
\end{figure}

To test stability, we add a local perturbation $\lambda \hat{V}=\lambda \sum_{j}\sigma_{j}^{x}$. The unperturbed case ($\lambda=0$) shows perfect revivals [Fig.~\ref{fig_xorX}(a)]. Under perturbation, we fit the decay envelope with a Gaussian; the extracted rate follows
a linear law $1/\tau=\kappa_1\lambda$ [Fig.~\ref{fig_xorX}(b)], consistent with the $\mathcal{O}(\lambda^{-1})$ bound. The inset shows the slope $\kappa_1$ approaching a finite value with increasing $L$.

\begin{figure}[!t]
  \begin{center}
  \includegraphics[width=\columnwidth]{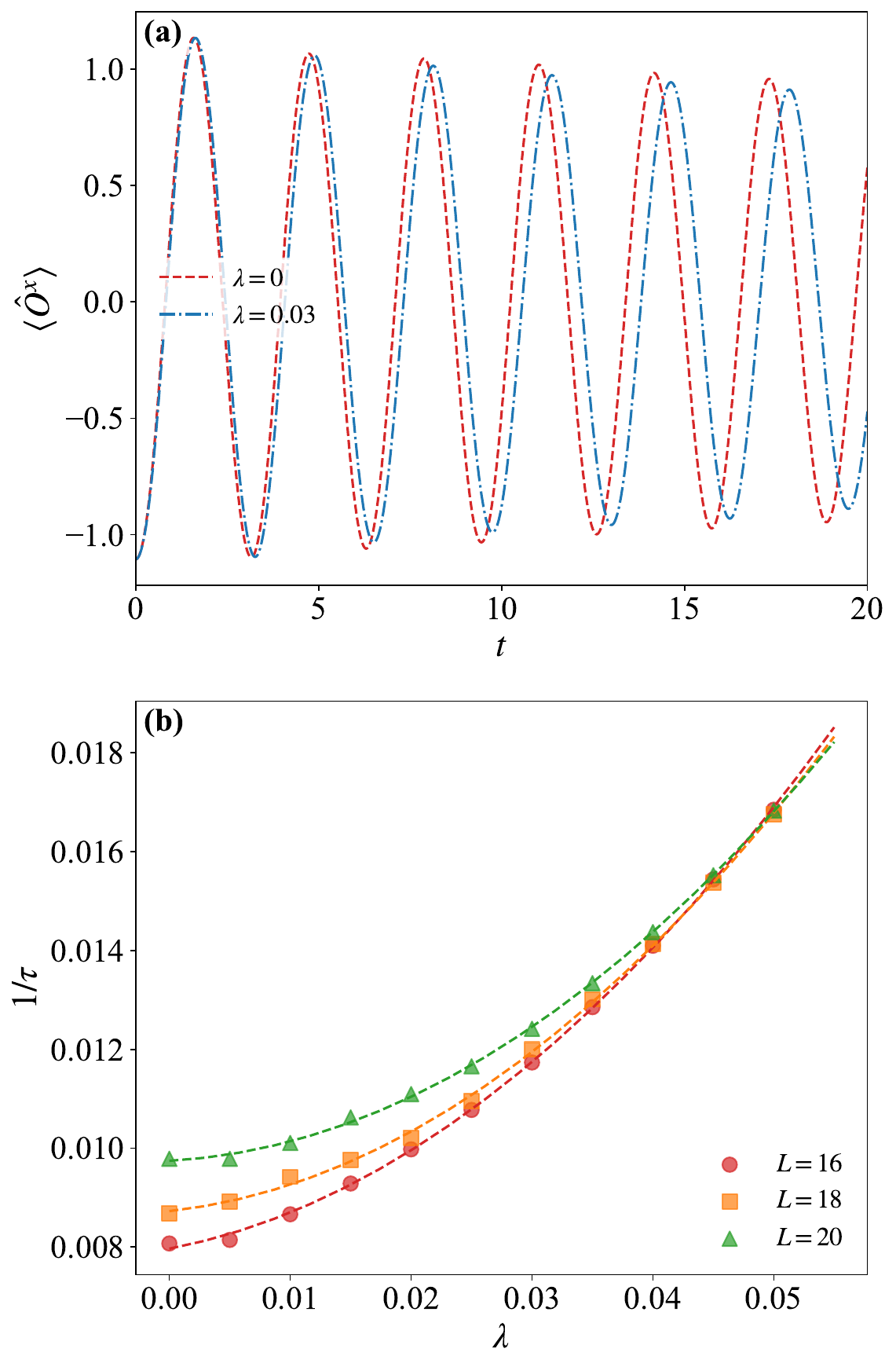}
  \end{center}
   \caption{\label{fig_deform_xorX} 
  (a) Dynamics of $\langle \hat{O}^{x} \rangle$ in the spin-1/2 deformed DWC model for $L=20$. Revivals are imperfect even at $\lambda=0$, indicating approximate QMBS.
   (b)  Thermalization rate $1/\tau$ (dots) as a function of $\lambda^2$ for various system sizes, with $\delta=0.03$. The decay rate $1/\tau$ is extracted by fitting the dynamic evolution with a Gaussian function. Dashed lines represent quadratic fits $1/\tau \propto  \lambda^2$.}
\end{figure}

Adding a transverse field term to the DWC model breaks domain wall conservation, transforming exact scars into approximate ones. The deformed DWC model is given by
\begin{equation}\label{ham:dDWC}
  \hat{H}_{\textrm{dDWC}} = \hat{H}_{\textrm{DWC}} + \delta \sum_{i=1}^{L} \sigma_{i}^{x},
\end{equation}
where $\delta$ is the strength of the additional transverse magnetic field. 
For any $\delta \neq 0$, the algebraic structure supporting exact scars is broken, and the scarring becomes approximate. 
For simplicity, we choose $\delta = 0.03$. As evidenced by the imperfect revivals in the unperturbed dynamics [Fig.\ref{fig_deform_xorX}(a)], the scar is indeed approximate. Upon applying an additional perturbation $\lambda \sum_{i} \sigma^{z}_{i}$,  the thermalization dynamics primarily exhibit a quadratic scaling, $1/\tau \propto \lambda^2$ [Fig.\ref{fig_deform_xorX}(b)], consistent with our observations in other approximate scar models. This controlled example thus reinforces our conclusion that the thermalization timescale sharply distinguishes between exact and approximate QMBS.

\section{Discussion and Conclusion} \label{sec4}
In this work, we have presented numerical and analytical evidence that the nonthermal features associated with quantum many-body scars (QMBS) persist for parametrically long times after a quench in generic closed systems subject to local perturbations of strength $\mathcal{O}(\lambda)$. Compared with Ref.~\cite{Lin2020}, our results establish a tighter lower bound on the thermalization time. For 1D models with exact scars, we prove a lower bound of $t^* \ge \kappa_1 \lambda^{-1}$ and consistently observe $\tau \propto \lambda^{-1}$ in numerics; for models with approximate scars, our numerics show a slower decay $\tau \propto \lambda^{-2}$. These conclusions apply when the initial state is a scar eigenstate and also for certain special product states, such as N\'eel states. We conjecture that the scaling laws we uncover persist in the thermodynamic limit. 
Our results are closely related to asymptotic quantum many-body scars~\cite{PhysRevLett.131.190401}, where tuning quasiparticle momenta toward $\pi$ causes the energy variance to vanish with increasing system size and the relaxation time to diverge.

Across the families we studied, the scaling in 1D systems organizes as follows. In the spin-1 XY model with both periodic boundary conditions (PBC) and open boundary conditions (OBC), the deformed PXP model with PBC, and the domain-wall-conserving (DWC) model with PBC, the thermalization time satisfies at least a $\mathcal{O}(\lambda^{-1})$ lower bound and numerically follows $\tau \propto \lambda^{-1}$. By contrast, in the spin-1 Kitaev model with PBC, the deformed PXP model with OBC, and the deformed DWC model with PBC, we find $\tau \propto \lambda^{-2}$. The latter scaling suggests that approximate scars are only weakly isolated from the thermal manifold—in the sense that generic perturbations induce finite matrix elements of order $\lambda$ to a dense set of ETH eigenstates—leading, via Fermi’s golden rule, to $\Gamma \simeq 2\pi |V|^2 \rho \propto \lambda^2$ and hence $\tau \sim \Gamma^{-1} \propto \lambda^{-2}$. For approximate scars, a promising route to sharpen bounds is to view quenched states as metastable and to adapt general arguments for metastability~\cite{PhysRevX.15.011064}. In this framework, the decay of local correlators can be nonperturbatively long in the inverse metastability radius, potentially extending beyond the golden-rule estimate. Developing a rigorous version of this argument for constrained and Rydberg-type scarred models is a natural direction for future work.

Our results suggest a counterintuitive organizing principle: exact scars can be less robust to generic perturbations than approximate scars. This fragility is expected when exact scarring relies on finely tuned algebraic structures, e.g., restricted-spectrum-generating algebras. Perturbations that disrupt such  structures immediately destroy the exact-eigenstate property~\cite{b83y-h128}. In a complementary classical picture, the periodic orbits underlying exact scars are structurally unstable under perturbations. Nevertheless, weak perturbations need not cause immediate thermalization; instead, they produce extended prethermal plateaus whose lifetime follows the $\lambda$-scaling identified above. 


The scaling can be benchmarked in 1D Rydberg-blockaded chains by quenching from the charge-density-wave (N\'{e}el) state and measuring revival envelopes, local correlators, or the Loschmidt amplitude while scanning a calibrated local perturbation of strength $\lambda$. 
A scaling analysis of $\tau(\lambda)$ then discriminates $\lambda^{-1}$ from $\lambda^{-2}$.
While ED provides a reliable approach in one dimension, it becomes impractical for higher-dimensional systems due to the exponential growth of the Hilbert space. Nevertheless, our conclusions naturally extend to higher-dimensional platforms, where experimental verification becomes feasible.
Two-dimensional (2D) Rydberg arrays now support high-fidelity entangling gates~\cite{PhysRevLett.114.100503} and flexible lattice engineering with optical tweezers~\cite{PhysRevX.4.021034}, enabling controlled tests of nonthermal dynamics in two dimensions.   
Building on proposals for quasiperiodic revivals as a scarring signature in 2D Rydberg-blockaded systems~\cite{PhysRevB.101.220304}, the same quench protocol can be used to extract $\tau(\lambda)$ within accessible windows.

The data that support the findings of this article are openly available~\cite{data}.

\begin{acknowledgments}
The authors appreciate very insightful discussions with Wen-Yi Zhang and Qing-Min Hu.
This work is supported by Postgraduate Research \& Practice Innovation Program of Jiangsu Province under No. KYCX25\_0640, the National Natural Science Foundation of
China (NSFC) under Grant No. 12174194 and stable support for basic institute research under Grant No. 190101.  
\end{acknowledgments}

\normalem
\bibliography{qmbs} 

\clearpage
\widetext
\setcounter{equation}{0}
\setcounter{figure}{0}
\setcounter{table}{0}
\setcounter{page}{1}
\setcounter{section}{0}
\setcounter{tocdepth}{0}
\renewcommand{\thefigure}{S\arabic{figure}}
\numberwithin{equation}{section} 
\begin{center}
{\bf \large Supplemental Material for “Tighter thermalization bounds for perturbed quantum many-body scars” }
\end{center}

\begin{center}
\large
Meng-Yun Mao${}^{1,2}$, Zhixiang Sun${}^{3,\ast}$, and Wen-Long You${}^{1,2,\dagger}$

\vspace{0.5em}
\normalsize
\textit{${}^{1}$College of Physics, Nanjing University of Aeronautics and Astronautics, Nanjing 211106, China} \\
\textit{${}^{2}$Key Laboratory of Aerospace Information Materials and Physics (NUAA), MIIT, Nanjing 211106, China} \\
\textit{${}^{3}$Center for Joint Quantum Studies and Department of Physics, School of Science, Tianjin University, 300072 Tianjin, China} \\

\vspace{1em}

\end{center}

 
\section{Derivation of Thermalization Time Bound} 
Consider a global quench scenario, where system evolves under a perturbed Hamiltonian:
\begin{equation}
\hat{H} = \hat{H}_{0} + \lambda \hat{V}, \quad | \psi(t) \rangle = e^{-i\hat{H} t} | \psi(0) \rangle,
\end{equation}
with the initial state $| \psi(0) \rangle$ being either an eigenstate of $\hat{H}_{0}$ or a superposition of a small number of its eigenstates. 
Both $\hat{H}_{0}$ and $\hat{V}$ are local Hamiltonians, with $\hat{V} = \sum_{j} \hat{v}_j$, where each $\hat{v}_j$ acts only on degrees of freedom near site $j$. We consider a local observable $\hat{m}_{0}$ localized near the origin, i.e., site $j = 0$.
   
\subsection{Revisiting Thermalization Time Bound $ t^* \sim \mathcal{O}(\lambda^{-1/2}) $} \label{review}
Given that the initial state $| \psi(0) \rangle$ is an eigenstate of the unperturbed Hamiltonian $\hat{H}_{0}$, the time derivative of the expectation value of the local observable $\hat{m}_{0}$ is given by 
\begin{eqnarray}
  \frac{d}{dt} \langle \psi(t) | \hat{m}_{0} |\psi(t) \rangle & = & i \langle \psi(t) | [\hat{H} , \hat{m}_{0}] | \psi(t) \rangle \nonumber \\
  & = & i \langle \psi(0) | [\hat{H} , e^{i \hat{H} t} \hat{m}_{0} e^{-i \hat{H} t}] | \psi(0) \rangle \nonumber \\
  & = & i \langle \psi(0) | [\lambda \hat{V} , e^{i \hat{H} t} \hat{m}_{0} e^{-i \hat{H} t}] | \psi(0) \rangle \nonumber,
\end{eqnarray}
where in the last line we  used the fact that $| \psi(0) \rangle$ is the eigenstate of $\hat{H}_{0}$.

A conservative estimate of the commutator norm can be obtained by noting that the Heisenberg-evolved operator $e^{i \hat{H} t} \hat{m}_{0} e^{-i \hat{H} t}$ is significantly supported only within the region  $|j| \leq v_{\rm LR} t$, where $v_{\rm LR}$ is the Lieb-Robinson velocity associated with  $\hat{H}$.
Therefore, the commutator with  $\hat{v}_{j}$ is appreciable only when $j$ lies within this effective light cone.
 
For sites inside the light cone  $|j| \leq v_{\rm LR} t$, we estimate the commutator norm using submultiplicativity and unitarity:
\begin{eqnarray} \label{scale}
  \parallel[\hat{v}_{j} , e^{i \hat{H} t} \hat{m}_{0} e^{-i \hat{H} t}]\parallel & \leq & 2 \parallel\hat{v}_{j}\parallel \parallel e^{i \hat{H} t} \hat{m}_{0} e^{-i \hat{H} t} \parallel = 2 \parallel \hat{v}_{j} \parallel \parallel\hat{m}_{0}\parallel \leq c_0',
\end{eqnarray}
where the last inequality defines a constant  $c_0'=\mathcal{O}(1)$, independent of system size.

For sites outside the light cone $|j| > v_{\rm LR} t$,  the Lieb-Robinson bound gives:
\begin{eqnarray} 
\label{Lieb-Robinson}
\parallel [\hat{v}_j , \hat{m}_{0}(t)] \parallel \leq c e^{-a(j - v_{\rm LR} t)},
 \end{eqnarray}
 which  leads to an exponentially suppressed contribution:
\begin{eqnarray} 
\left|\left\langle \psi(0) \left| \left[ \sum_{|j| > v_{\rm LR} t} \hat{v}_{j} , \hat{m}_{0}(t) \right] \right| \psi(0) \right\rangle \right| \le c_0'',
\end{eqnarray}
where $c_0''$ is a constant of order $\mathcal{O}(1)$.

Combining both regions, we obtain:
\begin{eqnarray} \label{lin_bound}
  & & | \langle \psi(0) | [\lambda \hat{V} , e^{i \hat{H} t} \hat{m}_{0} e^{-i \hat{H} t}] | \psi(0) \rangle | \nonumber \\
  & \leq & |  \langle \psi(0) | [\lambda \sum_{|j| \leq v_{\rm LR}t} \hat{v}_{j} , e^{i \hat{H} t} \hat{m}_{0} e^{-i \hat{H} t}] | \psi(0) \rangle |  +  | \langle \psi(0) | [\lambda \sum_{|j| > v_{\rm LR}t}\hat{v}_{j} , e^{i \hat{H} t} \hat{m}_{0} e^{-i \hat{H} t}] | \psi(0) \rangle | \nonumber \\
  & \leq & \lambda(c_0' v_{\rm LR} t + c_0'') \equiv \lambda(c_0 + c_1 t).
\end{eqnarray}
where $c_0$  and $c_1$ are constants of order $\mathcal{O}(1)$.

Integrating over time yields the bound on the deviation of the observable from its initial value:  
\begin{eqnarray}
  & & \left| \langle \psi(t) | \hat{m}_{0} | \psi(t) \rangle - \langle \psi(0) | \hat{m}_{0} | \psi(0) \rangle \right|  
  = \left| \int_{0}^{t} dt \frac{d}{dt} \langle \psi(t) | \hat{m}_{0} | \psi(t) \rangle \right|    \nonumber \\ 
  & \leq&   \int_{0}^{t} dt \big|\langle \psi(0) | [\lambda \hat{V} , e^{i \hat{H} t} \hat{m}_{0} e^{-i \hat{H} t}] | \psi(0) \rangle\big| \nonumber \\
  & \leq & \lambda (c_0 t +c_1 t^2 / 2).
\end{eqnarray}
It follows that the observable $\hat{m}_{0}$ remains close to its initial value up to a timescale~\cite{Lin2020}
\begin{equation}
    t^{*} = \mathcal{O}(\lambda^{-1/2}).
\end{equation}
In other words, thermalization does not occur before this characteristic time. In higher dimensions $d$, the volume of the light cone grows as 
$t^d$, leading to a general bound:
\begin{eqnarray}
t^{*} \sim \mathcal{O}(\lambda^{-1/(d+1)}).
\end{eqnarray}

However, the rough estimate used in Eq.~\eqref{scale} is overly conservative,  as it applies only to the special case where the initial state is a single eigenstate of $\hat{H}_{0}$ and neglects the detailed structure of both the Hamiltonian and the observable $\hat{m}_0$. 
Our approach addresses these limitations by systematically incorporating both the dynamical properties of the Hamiltonian and the structure of the observable.

\subsection{Derivation  of Eq.~\eqref{eq:deltam}} \label{AppA} 

The emergence of quantum many-body scars (QMBS)  can be understood through the framework of a spectrum generating algebra (SGA). An SGA provides an algebraic structure that generates towers of eigenstates with an equally spaced spectrum. For the specific case of QMBS, where only a subset of these towers might be preserved under perturbation, the more specific concept of a restricted spectrum generating algebra (RSGA)  is central~\cite{PhysRevB.101.195131, PhysRevB.102.085140}.

We assume the initial state is a superposition of eigenstates of $\hat{H}_0$ belonging to a scar subspace in a 1D system:
\begin{equation}
|\psi(0)\rangle = \sum_n c_n |\phi_n\rangle,
\end{equation}
where $\hat{H}_0 |\phi_n\rangle = E_n |\phi_n\rangle$ and $E_n = E_0 + n \mathcal{E}$ defines an equidistant spectrum of spacing $\mathcal{E} > 0$. The dimension of this subspace is finite, a key feature captured by the RSGA structure.

Considering a nonintegrable Hamiltonian $\hat{H}_0$ and a ladder operator $ \hat{Q}^\dagger$, along with a normalized reference state $|\phi_0\rangle$ satisfying 
\begin{eqnarray}
\hat{Q}^\dagger|\phi_0\rangle \neq 0, \quad \langle \phi_0 | \phi_0 \rangle = 1.
\end{eqnarray}
Crucially, we assume that $|\phi_0\rangle$ is a  “lowest-weight” state with respect to the ladder operator, meaning it cannot be generated by the action of $\hat{Q}^\dagger$ on any other state in the Hilbert space.
Under these assumptions, the Hamiltonian $\hat{H}_0$ is said to exhibit a {\it restricted spectrum generating algebra of order $\zeta$} (RSGA-$\zeta$) if the following conditions are satisfied:
\begin{align}
\hat{H}_0|\phi_0\rangle &= E_0 |\phi_0\rangle, \label{eq:rsga1} \\
[\hat{H}_0, \hat{Q}^\dagger]|\phi_0\rangle &= \mathcal{E} \hat{Q}^\dagger|\phi_0\rangle, \label{eq:rsga2} \\
\underbrace{[[[\hat{H}_0, \hat{Q}^\dagger], \hat{Q}^\dagger], \cdots, \hat{Q}^\dagger]}_{n\text{ times}} |\phi_0\rangle &= 0,\quad \text{for } 2 \le n \le \zeta, \label{eq:rsga3} \\
\underbrace{[[[\hat{H}_0, \hat{Q}^\dagger], \hat{Q}^\dagger], \cdots, \hat{Q}^\dagger]}_{\zeta+1\text{ times}} &= 0,\quad \text{for } n =\zeta+1, \label{eq:rsga4}
\end{align}
where $\zeta$ is the nested-commutator closure depth. All models studied in this work satisfy the RSGA-1 condition, ensuring that the repeated action of $ \hat{Q}^\dagger$ on $|\phi_0\rangle$ generates a finite tower of $\mathcal{L}+1$ linearly independent states:
\begin{equation}
|\tilde{\phi}_n\rangle = (\hat{Q}^\dagger)^n |\phi_0\rangle, \quad \text{for } 0 \le n \le \mathcal{L},
\end{equation}
with the truncation condition $(\hat{Q}^{\dagger})^{\mathcal{L}+1} | \phi_0 \rangle=0$.
Here, $\mathcal{L}$ denotes the dimension of nontrivial (i.e., excited) scar states in the tower and typically scales linearly with the system size, i.e., $\mathcal{L} = \mathcal{O}(L)$.
These unnormalized QMBS  satisfy the eigenvalue relation
\begin{equation}
\hat{H}_0 |\tilde{\phi}_n\rangle = (E_0 + n\mathcal{E}) |\tilde{\phi}_n\rangle. \label{eq:eigenstate_qmbs}
\end{equation}
The corresponding normalized scarred states are defined as
\begin{equation}
|\phi_n\rangle \equiv \frac{|\tilde{\phi}_n\rangle}{\sqrt{\langle \tilde{\phi}_n | \tilde{\phi}_n \rangle}}.
\end{equation}

In the Schr\"{o}dinger picture, the time evolution of a state is generated by the unitary operator 
\begin{equation}
    U(t) = e^{-i \hat{H} t},
\end{equation} such that the time-evolved state is given by  $| \psi(t) \rangle = U(t) | \psi(0) \rangle$.
The corresponding state in the interaction picture is defined as
\begin{equation}
  | \psi_{\textrm{I}}(t) \rangle = e^{i \hat{H}_{0} t} | \psi(t) \rangle \equiv U_{0}^{\dagger}(t) | \psi(t) \rangle.
\end{equation}
where $U_{0}(t)=e^{-i \hat{H}_{0} t}$ denotes the unperturbed evolution operator. In this picture, operators evolve as  $\hat{m}_{\textrm{I}}(t) = U_{0}^{\dagger}(t) \hat{m}_{0} U_{0}(t)$, $\hat{V}_{\textrm{I}}(t) = U_{0}^{\dagger}(t) \hat{V} U_{0}(t)$ and $(\hat{v}_j)_{\textrm{I}}(t) = U_{0}^{\dagger}(t) \hat{v_j} U_{0}(t)$.
The full time-evolution operator can be decomposed as
\begin{equation}
  U(t) = U_{0}(t) U_{\textrm{I}}(t),
\end{equation}
where the interaction-picture evolution operator $ U_{\textrm{I}}(t)$ is given by the time-ordered exponential
\begin{equation}
  U_{\textrm{I}}(t) = \mathcal{T} \exp(- i \lambda \int_{0}^{t} dt' \hat{V}_{\textrm{I}}(t')).
\end{equation}

Using this decomposition, the operator $\hat{m}_0$ in the Heisenberg picture can be expressed as 
\begin{equation}
  \label{eq:m}
  e^{i \hat{H} t} \hat{m}_{0} e^{-i \hat{H} t} = U^{\dagger}(t) \hat{m}_{0} U(t) = U_{\textrm{I}}^{\dagger}(t) \hat{m}_{\textrm{I}}(t) U_{\textrm{I}}(t).
\end{equation}
Expanding  $U_{\textrm{I}}(t)$ to first order in $\lambda$, we have
\begin{equation}
  U_{\textrm{I}}(t) \approx 1 - i \lambda \int_{0}^{t} dt' \hat{V}_{\textrm{I}}(t') + \mathcal{O}(\lambda^{2}).
\end{equation}
Substituting this into Eq.~\eqref{eq:m}, we obtain
\begin{eqnarray}
    e^{i \hat{H} t} \hat{m}_{0} e^{-i \hat{H} t} & = & \hat{m}_{\textrm{I}}(t) + i \lambda \int_{0}^{t} dt' \hat{V}_{\textrm{I}}(t') \hat{m}_{\textrm{I}}(t) - i \lambda \int_{0}^{t} dt' \hat{m}_{\textrm{I}}(t) \hat{V}_{\textrm{I}}(t') + \mathcal{O}(\lambda^{2}) \nonumber\\
    & = & e^{i \hat{H}_0 t} \hat{m}_{0} e^{-i \hat{H}_0 t} + i \lambda \int_{0}^{t} dt' [\hat{V}_{\textrm{I}}(t') , \hat{m}_{\textrm{I}}(t)] + \mathcal{O}(\lambda^{2}). \label{eq:tilde_m_expansion}
\end{eqnarray}
Then, it can be straightforwardly shown that the deviation [Eq.~(2)] in the expectation value of $\hat{m}_{0}$ between the perturbed and unperturbed evolutions at time $t$ is given by
\begin{equation}
    \delta m_0 (t) = i \lambda \sum_{j} \int_{0}^{t} dt' \langle \psi(0) | [(\hat{v}_j)_{\textrm{I}}(t') , \hat{m}_{\textrm{I}}(t)] | \psi(0) \rangle + \mathcal{O}(\lambda^{2}). \label{eq:Vm}
\end{equation}

For sites satisfying $|j| \leq v_{\rm LR} t$, we can get that
\begin{equation}
    \label{eq:bound}
    \left| \int_{0}^{t} dt' \langle \psi(0) | [(\hat{v}_j)_{\textrm{I}}(t') , \hat{m}_{\textrm{I}}(t)] | \psi(0) \rangle \right| \leq c_1'.
\end{equation}
This estimate relies on the structure of the initial state and the spectrum of the unperturbed Hamiltonian.

To be more precise, we denote the integrand as
\begin{equation}
    f(t', t) \equiv \langle \psi(0)|[(\hat{v}_j)_{\mathrm{I}}(t'), \hat{m}_{\mathrm{I}}(t)] | \psi(0) \rangle.
\end{equation}
We can expand $f(t', t)$ in terms of the eigenstates $|\phi_n\rangle$ of $\hat{H}_0$ as follows:
\begin{equation}
    f(t', t) = \sum_{a,b} c_a c_b \mathcal{M}_{ab} (t' , t),
\end{equation}
where
\begin{equation}
    \mathcal{M}_{ab} (t' , t) = \sum_{c} \left( e^{i (E_a - E_c) t'} e^{i (E_c - E_b) t}  \langle \phi_a | \hat{v}_j | \phi_c \rangle \langle \phi_c | \hat{m}_{0} | \phi_b \rangle - e^{i (E_a - E_c) t} e^{i (E_c - E_b) t'} \langle \phi_a | \hat{m}_{0} | \phi_c \rangle \langle \phi_c | \hat{v}_j | \phi_b \rangle \right).
\end{equation}
In short,
\begin{equation}
    \mathcal{M}_{ab} (t' , t) = \sum_{c} \left( e^{i (E_a - E_c) t'} e^{i (E_c - E_b) t} (v_j)_{ac} (m_0)_{cb} - e^{i (E_a - E_c) t} e^{i (E_c - E_b) t'} (m_0)_{ac} (v_j)_{cb} \right).
\end{equation}
Note that the sums over $a,b$ are finite (over the scar states), and the sums over $c$ are over the entire Hilbert space. 
It thus can be yielded that
\begin{eqnarray}
    \int_{0}^{t} dt' f (t' , t) & = & \sum_{a, b} c_a c_b \sum_{c} \left( \frac{e^{i (E_a - E_b) t} - e^{i (E_c - E_b) t}}{i (E_a - E_c)} (v_j)_{ac} (m_0)_{cb} - \frac{e^{i (E_a - E_b) t} - e^{i (E_a - E_c) t}}{i (E_c - E_b)} (m_0)_{ac} (v_j)_{cb} \right) \nonumber \\
    & = & \sum_{a, b} c_a c_b \sum_{c} \left( \frac{2 (v_j)_{ac} (m_0)_{cb}}{i (E_a - E_c)} \cos \left(E_a-E_b\right)t - \frac{2 (v_j)_{ac} (m_0)_{cb}}{i (E_a - E_c)} \cos \left(E_c-E_b\right)t \right) \nonumber \\
    & = & \sum_{a, b , c} c_a c_b \frac{4 (v_j)_{ac} (m_0)_{cb}}{i (E_a - E_c)} \sin \left(\frac{(2E_b - E_a - E_c) t}{2}\right) \sin \left(\frac{ (E_a - E_c) t}{2}\right).  \label{eq:integrand}
\end{eqnarray}
At first sight, the prefactor $1/(E_a - E_c)$ in Eq.~\eqref{eq:integrand} appears to lead to a divergence when $E_a \to E_c$, for instance when a scar state is nearly degenerate with a thermal state.  
However, this apparent singularity is in fact removable: the factor $\sin[(E_a - E_c)t/2]$ in the numerator vanishes linearly in $(E_a - E_c)$ and cancels the would-be pole. As a consequence, the time integral of $f(t',t)$ does not generate any secular (linearly growing) contribution. Using the inequality $|\sin x|\le 1$, we obtain the time-independent bound
\begin{equation}
  \label{eq:bound_f}
  \left| \int_{0}^{t} dt' f (t' , t) \right| \leq \sum_{a, b , c, E_c \not= E_a} \left|c_a c_b \frac{4 (v_j)_{ac} (m_0)_{cb}}{E_a - E_c} \right|.
\end{equation}
To estimate the right-hand side of Eq.~\eqref{eq:bound_f}, we invoke the eigenstate thermalization hypothesis (ETH) for local operators. For a generic thermal eigenstate basis $\{|\phi_\alpha\rangle\}$ and a local operator $\hat{O}$, ETH states that
\begin{equation}
\label{eq:ETH}
\langle \phi_\alpha | \hat{O} | \phi_\beta \rangle
= O(\bar E)\,\delta_{\alpha\beta}
+ e^{-S(\bar E)/2}\, f_O(\bar E,\omega)\, R_{\alpha\beta},
\end{equation}
where $\bar E = (E_\alpha+E_\beta)/2$, $\omega = E_\alpha - E_\beta$, $S(\bar E)$ is the thermodynamic entropy at energy $\bar E$, $f_O$ is a smooth function, and $R_{\alpha\beta}$ is a random variable with zero mean and unit variance. Equation~\eqref{eq:ETH} implies that the off-diagonal matrix elements of local operators between typical thermal eigenstates are exponentially suppressed in the system size, scaling as $\sim e^{-S(\bar E)/2}$.
In our case, both $\hat{v}_j$ and $\hat{m}_0$ are local operators. Thus, for generic thermal eigenstates, their off–diagonal matrix elements $(v_j)_{ac}$ and $(m_0)_{cb}$ are exponentially small in system size, while scar eigenstates overlap significantly only with a parametrically small subset of thermal states. Consequently, only a limited number of terms in the triple sum in Eq.~\eqref{eq:bound_f} contribute appreciably. Consistent with this ETH-based expectation, our numerical data show that the full sum remains of order $\mathcal{O}(1)$ as the system size increases. We can therefore introduce a constant $c_1' \sim \mathcal{O}(1)$, independent of both $t$ and $L$, such that
\begin{equation}
\left| \int_{0}^{t} dt'\, f(t' , t) \right| \le c_1',
\end{equation}
which is precisely the bound used in Eq.~\eqref{eq:bound}.

For $|j| > v_{\rm LR} t$, the Lieb-Robinson bound gives:
\begin{eqnarray} 
\label{Lieb-Robinson1}
\parallel [\hat{v}_j , \hat{m}_{0}(t)] \parallel \leq c e^{-a(j - v_{\rm LR} t)}.
\end{eqnarray}
Since the bound is invariant under time translations and under unitary transformations of the operators, we can extend it to operators in the interaction picture evaluated at two arbitrary times $t$ and $t'$. In particular, we obtain


\begin{equation}
    \parallel [(\hat{v}_j)_{\rm I}(t') , \hat{m}_{\rm I}(t)] \parallel \leq c e^{-a(j - v_{\rm LR} (t - t'))}.
\end{equation}
We now use this bound to control the time integral of the commutator, given by
\begin{eqnarray}
    & & \left| \int_{0}^{t} dt' \langle \psi(0) | [(\hat{v}_j)_{\textrm{I}}(t') , \hat{m}_{\textrm{I}}(t)] | \psi(0) \rangle \right| \leq  \int_{0}^{t} dt' \parallel [(\hat{v}_j)_{\textrm{I}}(t') , \hat{m}_{\textrm{I}}(t)] \parallel \nonumber \\
    & \leq & \int_{0}^{t} dt' c e^{-a(j - v_{\rm LR} (t - t'))} \leq c e^{-a(j - v_{\rm LR} t)} \int_{0}^{t} dt' e^{-a v_{\rm LR} t'} \leq \frac{c}{a v_{\rm LR}} e^{-a(j - v_{\rm LR} t)}.
\end{eqnarray}
Summing over all sites outside the light cone yields
\begin{equation}
   \sum_{|j| > v_{\rm LR} t} \left| \int_{0}^{t} dt' \langle \psi(0) | [(\hat{v}_j)_{\textrm{I}}(t') , \hat{m}_{\textrm{I}}(t)] | \psi(0) \rangle \right| \leq \sum_{|j| > v_{\rm LR} t} \frac{c}{a v_{\rm LR}} e^{-a(j - v_{\rm LR} t)} \leq c_1''.
\end{equation}
Thus, we obtain the bound
\begin{eqnarray}
   | \delta m_0(t) | & \leq & \lambda \sum_{j} \left| \int_{0}^{t} dt' \langle \psi(0) | [(\hat{v}_j)_{\textrm{I}}(t') , \hat{m}_{\textrm{I}}(t)] | \psi(0) \rangle \right| \nonumber \\
   & = & \lambda \left( \sum_{|j| \leq v_{\rm LR} t} \left| \int_{0}^{t} dt' \langle \psi(0) | [(\hat{v}_j)_{\textrm{I}}(t') , \hat{m}_{\textrm{I}}(t)] | \psi(0) \rangle \right| + \sum_{|j| > v_{\rm LR} t} \left| \int_{0}^{t} dt' \langle \psi(0) | [(\hat{v}_j)_{\textrm{I}}(t') , \hat{m}_{\textrm{I}}(t)] | \psi(0) \rangle \right| \right)\nonumber \\
   & \leq & \lambda (v_{\rm LR} t \cdot c_1' + c_1'') \equiv \lambda (c_0 + c_1 t),
\end{eqnarray}
where $c_0$  and $c_1$ are constants of order $\mathcal{O}(1)$.

In higher dimensions $d$, the number of scar states scales as $\mathcal{L} \sim \mathcal{O}(L^d)$, where $L$ denotes the linear system size---that is, the number of lattice sites along each spatial direction in a $d$-dimensional hypercubic lattice.
The lieb-robinson bound can be written as
\begin{equation}
    \parallel [\hat{v}_{\bf x} , \hat{m}_{\bf y}(t)] \parallel \leq c e^{-a(|\textbf{x} - \textbf{y}| - v_{\rm LR} t)},
\end{equation}
where $\textbf{x}$ and $\textbf{y}$ denote the positions of local operators and \(|\mathbf{x} - \mathbf{y}|\) is the Euclidean distance between the two positions.
In a $d$-dimensional lattice, the number of sites $\mathbf{x}$ satisfying $|\mathbf{x} - \mathbf{y}| < v_{\rm LR} t$ then scales as the volume of a $d$-dimensional ball, i.e., as $\mathcal{O}\left((v_{\rm LR} t)^{d}\right)$.
Analogous to the one-dimensional derivation, the bound on the $|\delta m_0 (t)|$ generalizes to higher dimensions as
\begin{equation}
    | \delta m_0(t) | \leq \lambda (c_0 + c_1 t^{d}).
\end{equation}

\subsection{Lieb-Robinson-Type Bound and Operator Spreading in Scarred Systems}
 \label{OTOC}
The propagation of local perturbations can be understood through the lens of quantum information dynamics.
In generic closed quantum systems with local interactions, there exists a finite bound on the speed of information propagation—known as the Lieb-Robinson bound~\cite{cheneau_light-cone-like_2012}.  While the Lieb-Robinson velocity $v_{\rm LR}$ provides a state-independent upper bound on operator spreading, $v_{\rm B}$ acts as an effective, state-dependent velocity capturing the actual dynamics in specific systems. 
 In strongly correlated systems, the butterfly effect implies that initially local operators spread ballistically over time—an effect reflected in $v_{\rm B}$~\cite{PhysRevLett.117.091602}. 

To investigate this behavior in systems hosting QMBS, we use the out-of-time-ordered correlator (OTOC) as a diagnostic. The OTOC measures the non-commutativity of two initially local operators at different times, and it captures essential features of quantum information spreading and scrambling~\cite{maldacena_bound_2016, hashimoto_out--time-order_2017}. It is defined as: 
 \begin{equation}
\label{Ft}
  F_{ij}(t) = \langle \psi(0) |\hat{W}_{i}^{\dagger} \hat{V}_{j}^{\dagger}(t) \hat{W}_{i} V_{j}(t) | \psi(0) \rangle,
\end{equation}
where $|\psi(0)\rangle$ is an initial pure state, $\hat{W}_i$, $\hat{V}_j$ are local {\it unitary} observables defined on sites $i$, $j$, respectively.
The time-evolved operator is given by $\hat{V}_j(t) = e^{i \hat{H} t} \hat{V}_j e^{-i \hat{H} t}$.
A useful interpretation is that when the support of the Heisenberg-evolved operator $\hat{V}_j(t)$ has not yet reached site $i$, the operators commute, i.e., $[\hat{W}_i, \hat{V}_j(t)] = 0$, and thus $F_{ij}(t) = 1$. Once the support of $\hat{V}_j(t)$ overlaps with site $i$, this commutation breaks down and $F_{ij}(t) \ne 1$, signaling the onset of correlations between the two sites and the growth of operator complexity.

\begin{figure*}[!t]
  \begin{center}
  \includegraphics[width=0.7\textwidth]{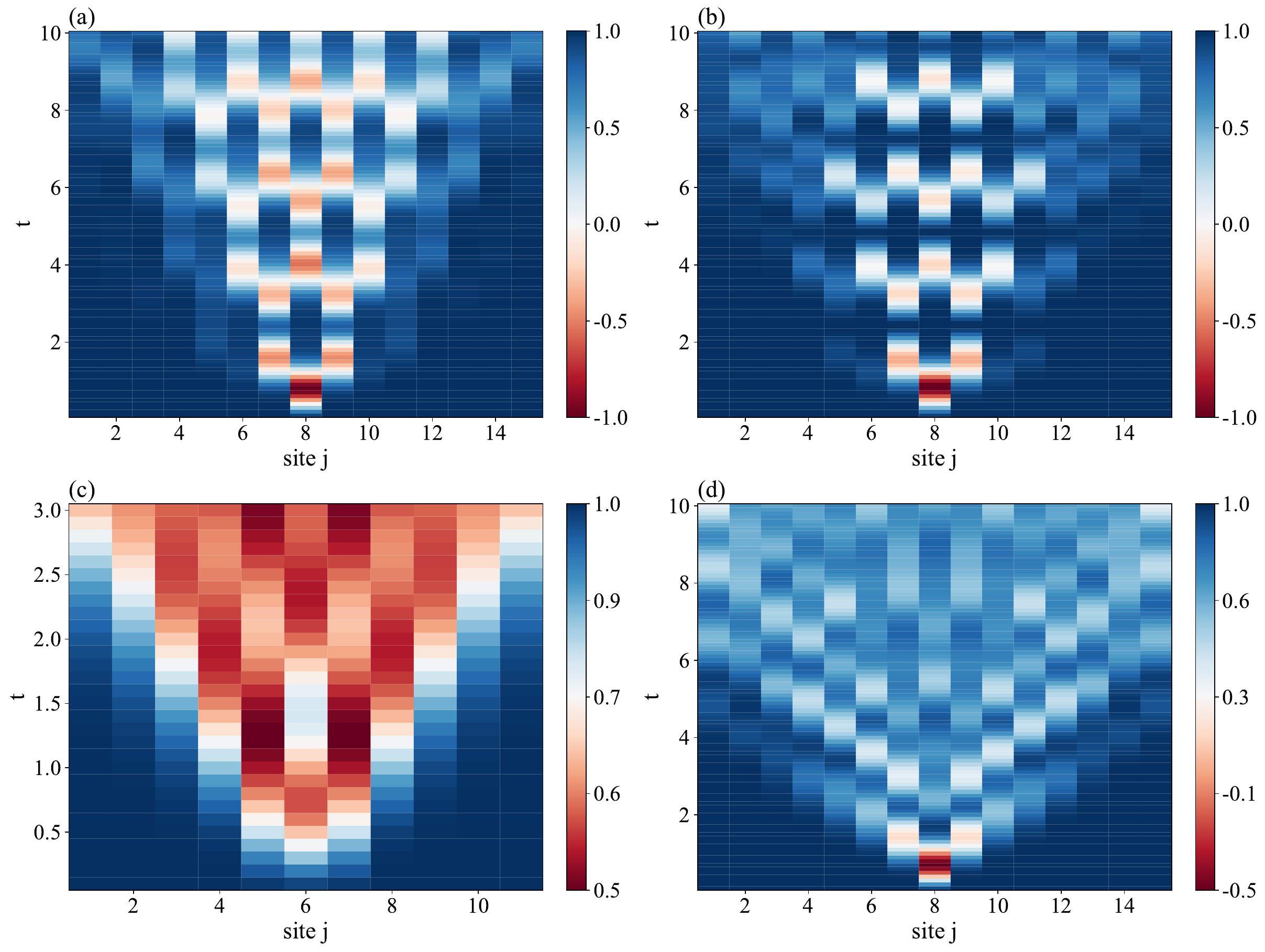}
  \end{center}
   \caption{\label{fig_otoc} Information scrambling dynamics in quantum many-body scarred systems.
Spatiotemporal evolution of the ZZ-OTOCs $F_{ij}(t)$ (\ref{Ft}) is shown for various models and initial states under open boundary conditions (OBC):
(a) PXP model with the N\'eel initial state $|\mathbb{Z}_2\rangle = |\downarrow\uparrow\downarrow \uparrow\cdots \downarrow\rangle$ for $L = 15$ and probe site $i = 8$;
(b) Deformed PXP model with the same N\'eel initial state, system size, and probe site;
(c) Spin-1 XY model with the nematic N\'eel initial state for $L = 11$ and $i = 6$;
(d) PXP model with the polarized initial state $|\mathbb{Z}_1\rangle = |\downarrow\downarrow\downarrow\cdots \downarrow\rangle$ for $L = 15$ and $i = 8$.}
\end{figure*}

Here we employ the ZZ-OTOC ($\hat{W}_i = \sigma_{i}^{z}$, $\hat{V}_j = \sigma_{j}^{z}$)  to diagnose operator spreading in the scarred dynamics. 
Figure~\ref{fig_otoc} shows results for: (a) the PXP model, (b) the deformed PXP model, and (c) the spin-1 XY model. The initial states are chosen as (exact or approximate) superpositions of scar eigenstates.
In all three cases, the OTOC exhibits ballistic spreading and forms a clear linear light-cone structure.
Inside the light cone, periodic revivals of the commutator reveal coherent, non-ergodic evolution characteristic of scarred dynamics.

In contrast, Fig.~\ref{fig_otoc}(d) shows the result for a non-scar initial state: the fully polarized state $|\mathbb{Z}_ 1\rangle = |\downarrow\downarrow\cdots\downarrow\rangle$ in the PXP model.
Here, the OTOC front has a shallower  slope comparing that for $|\mathbb{Z}_ 1\rangle$ state and the corresponding butterfly velocity is larger ($v_{\rm B} \sim 1$). 
Meanwhile,  oscillations within the light cone are negligible.
This behavior indicates faster and more chaotic information scrambling, with OTOCs decaying quickly and lacking revivals.



In other words, while $v_{\rm LR}$ sets a theoretical upper limit, it is primarily governed by thermal eigenstates and is therefore not a tight bound for QMBS. For scarred systems, a more pertinent constraint is the butterfly velocity $v_{\rm B}$ evaluated within the scar subspace, which reflects their characteristically slower and more structured operator spreading. In this work, we make the simplifying assumption that $v_{\rm LR}$ is constant. However, elucidating the precise relationship between the global Lieb-Robinson velocity and the suppressed scrambling rates within the scar sector remains an important open question.

\section{Spin-1 XY model}

To quantitatively assess the approximate linear scaling behavior, we decompose the thermalization rate $1/\tau$ into its fitted linear ($a_1\lambda$) and quadratic ($a_2\lambda^2$) components. Table~\ref{table_xy} summarizes these contributions for two representative perturbation strengths, $\lambda = 0.01$ and $0.05$, across different system sizes and boundary conditions.
In the weak-perturbation regime (e.g., $\lambda = 0.01$), the linear term $a_1\lambda$ is approximately one to two orders of magnitude larger than the quadratic correction $a_2\lambda^2$. This comparison confirms that the thermalization process is largely governed by leading-order perturbative processes at small $\lambda$. As the coupling strength increases to $\lambda = 0.05$, the quadratic contribution grows progressively, accounting for the subtle curvature observed in the numerical data. However, even at this relatively large $\lambda$, the linear term remains the primary contributor to the total rate. Taken together, these observations suggest that focusing on the linear response of $1/\tau$ to $\lambda$ captures the main trend of the data, while the higher-order terms provide subleading corrections.

\begin{table*}[t] 
  \renewcommand{\arraystretch}{1.3} %
  \caption{\textbf{Quadratic fitting coefficients for the spin-1 XY model.}
  We fit the thermalization rate as $1/\tau \approx a_1\lambda + a_2\lambda^2$ under PBC and OBC for system sizes $L=8$ and $L=12$.
  The table reports the contributions $a_1\lambda$ and $a_2\lambda^2$ evaluated at two representative perturbation strengths.}  
  \label{table_xy}
  \centering
  {\setlength{\tabcolsep}{12pt}
  \begin{tabular}{c|cc|cc}
      \hline\hline
  \multirow{2}{*}{\textbf{Case}} & \multicolumn{2}{c|}{$\lambda = 0.01$} & \multicolumn{2}{c}{$\lambda = 0.05$} \\  \cline{2-5}
  & $a_1 \lambda$ & $a_2 \lambda^2$ & $a_1 \lambda$ & $a_2 \lambda^2$ \\
      \hline
    L=8 (PBC) & 0.0096 & 0.0004 & 0.0478 & 0.0103 \\
    L=8 (OBC) & 0.0093 & 0.0001 & 0.0465 & 0.0032 \\
    L=12 (PBC) & 0.0078 & 0.0005 & 0.0388 & 0.0120 \\
    L=12 (OBC) & 0.0077 & 0.0003 & 0.0386 & 0.0065 \\
    \hline\hline
  \end{tabular}}
\end{table*}

\section{The deformed PXP model} \label{AppC}
Then, we replace  PBC with OBC to investigate boundary effects.
The Hamiltonian will be changed to 
\begin{equation}
  \hat{H}_{\rm dPXP} = \sum_{i = 2}^{L-1} P_{i-1} \sigma_{i}^{x} P_{i+1} + \sigma_{1}^{x} P_{2} + P_{L-1} \sigma_{L}^{x} +  \hat{H}_\delta ,
\end{equation}
where 
\begin{eqnarray}
\hat{H}_\delta  & = & - \sum_{\delta=2}^{R} \bigg(\sum_{i=2}^{L-\delta} P_{i-1} \sigma^{x}_{i} P_{i+1} \sigma^{z}_{i+\delta}   + \sum_{i=\delta+1}^{L-1} \sigma^{z}_{i-\delta} P_{i-1} \sigma^{x}_{i} P_{i+1} \bigg)   - \sum_{\delta=2}^{R} (\sigma^{x}_{1} P_{2} \sigma^{z}_{1+\delta} + \sigma^{z}_{L-\delta} P_{L-1} \sigma^{x}_{L}).
\end{eqnarray}
According to the evolution of the expectation value $\langle \hat{O}_{z} \rangle$ in the inset of Fig.~3,
we find that OBC makes the revival not perfect.
We also plot the thermalization time with respect to the perturbation strength in Fig.~3.
It can be seen that $1/\tau$ depends linearly on $\lambda$ under PBC, while it follows a quadratic trend under OBC.
Essentially, this behavior arises from the fact that the system exhibits a exact scar under PBC, whereas it manifests as a approximate scar under OBC. 
This simulation also supports the conclusions drawn from the previous two examples.

\begin{figure}[!t]
  \begin{center}
  \includegraphics[width=0.6\columnwidth]{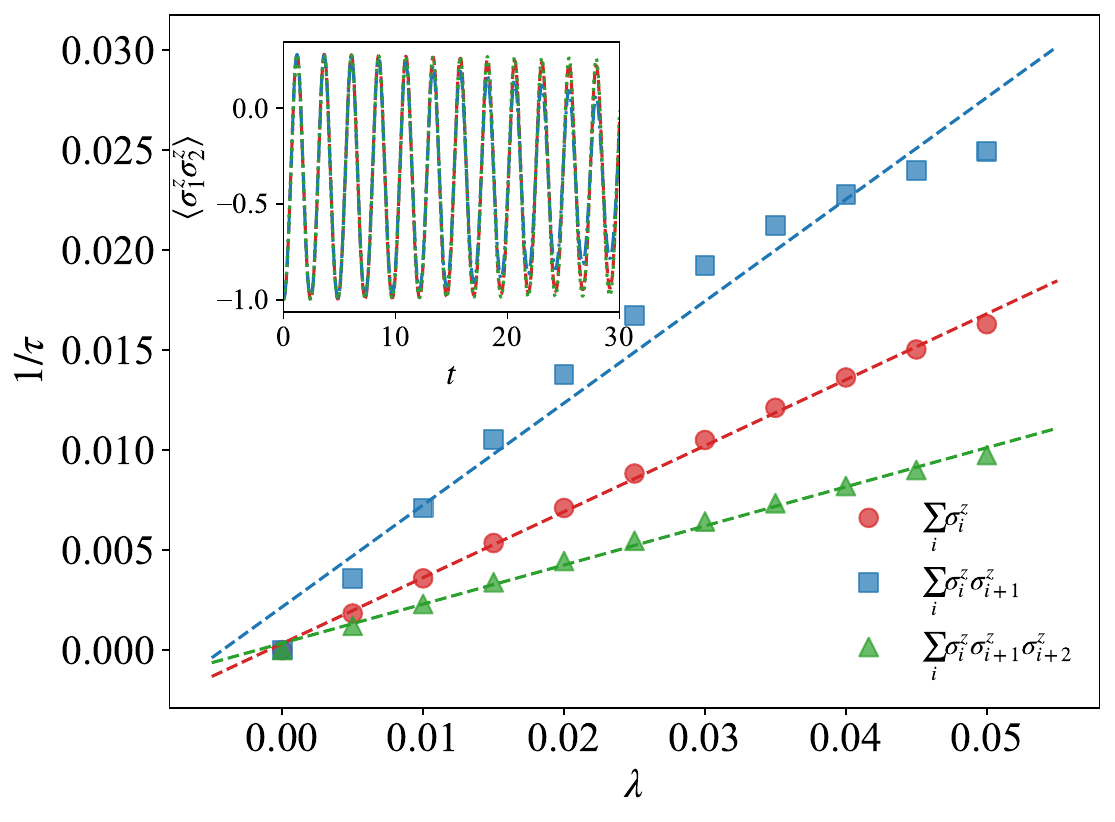}
  \end{center}
   \caption{\label{fig_operator} The dependence of the fitted decay time $\tau$ of the observable $\braket{\hat{O^z}}$ in deformed PXP model on perturbation strength for $V = \sum_{i} \sigma_{i}^{z}$, $V = \sum_{i} \sigma_{i}^{z} \sigma_{i+1}^{z}$ and $V = \sum_{i} \sigma_{i}^{z} \sigma_{i+1}^{z} \sigma_{i+2}^{z}$, respectively. 
   The system size is $L=24$.
   The inset shows the dynamics behavior of $\langle \sigma_{1}^{z} \sigma_{2}^{z} \rangle$ for different perturbation forms.}
\end{figure}

We next continue exploring the impact of different perturbation forms.
In Fig.~\ref{fig_operator}, we show the fitting of $\tau^{-1}$ and $\lambda$ for different $k$-local perturbations $\delta V = \sum_i \sigma_i^z$, $\delta V=\lambda\sum_i \sigma^z_i \sigma^z_{i+1}$ and $\delta V=\lambda\sum_i \sigma^z_i \sigma^z_{i+1} \sigma^z_{i+2}$ in the deformed PXP model. 
In the inset, we show the oscillations of two-site correlation $\langle \sigma_{1}^{z} \sigma_{2}^{z} \rangle$ for different perturbations.
We find that the dependence of the thermalization time on the perturbation strength is immune to changes in perturbation forms.
For large $\lambda$, due to the fast decay of the observable, there are deviations from the $\lambda^{-1}$ behavior.

\section{$t^{*} \sim \mathcal{O}(\lambda^{-2})$ scaling From Fermi's golden rule}
Consider a Hamiltonian $\hat{H}_{0}$ perturbed by time-dependent term $\hat{V}(t)$ with small perturbation strength $\lambda$.
The full Hamiltonian is given by $\hat{H} = \hat{H}_{0} + \lambda \hat{V}$. 
$|\phi_{n}\rangle$ is an eigenstates of $\hat{H}_{0}$ with the corresponding eigenstates $E_{n}$.
The time-evolved state can be written as $|\psi(t)\rangle = e^{-i\hat{H} (t - t_0)} |\psi(t_0)\rangle$.
In the interaction picture, the state is transformed to $|\psi_{\rm I}(t)\rangle = e^{i\hat{H}_{0} t} |\psi(t) \rangle$ and the Schr\"{o}dinger equation is described as 
\begin{equation}
  i \frac{\partial}{\partial t} | \psi_{\rm I}(t) \rangle = [e^{i \hat{H}_{0} t} \lambda \hat{V}(t) e^{-i \hat{H}_{0} t}] | \psi_{\rm I}(t) \rangle \equiv \lambda \hat{V}_{\rm I}(t) | \psi_{\rm I}(t) \rangle.
\end{equation}
Formally, the state vector at time $t$ in the interaction representation is obtained by integrating both sides:
\begin{equation}
  | \psi_{\rm I}(t) \rangle = | \psi_{\rm I}(t_0) \rangle - i \lambda \int_{t_0}^{t} dt' \hat{V}_{\rm I}(t') | \psi_{\rm I}(t') \rangle.
\end{equation}
Due to the appearance of the unknown state $| \psi_{\rm I}(t) \rangle$, we need to iterate the formal solution once:
\begin{equation}
  | \psi_{\rm I}(t) \rangle = | \psi_{\rm I}(t_0) \rangle - i \lambda \int_{t_0}^{t} dt' \hat{V}_{\rm I}(t') \bigg[ | \psi_{\rm I}(t_0) \rangle - i \lambda \int_{t_0}^{t'} dt'' \hat{V}_{\rm I}(t'') | \psi_{\rm I}(t'') \rangle \bigg],
\end{equation}
and then keep going:
\begin{equation}
  | \psi_{\rm I}(t) \rangle = | \psi_{\rm I}(t_0) \rangle - i \lambda \int_{t_0}^{t} dt' \hat{V}_{\rm I}(t') | \psi_{\rm I}(t_0) \rangle + (-i \lambda)^2 \int_{t_0}^{t} dt' \hat{V}_{\rm I}(t') \int_{t_0}^{t'} dt'' \hat{V}_{\rm I}(t'') | \psi_{\rm I}(t_0) \rangle + \cdots
\end{equation}
We thus obtain a formal perturbation series to many orders.
The series converges rapidly when the perturbation strength $\lambda$ is small.

Let $|\psi_{\rm I}(t_0)\rangle = |\phi_{0}\rangle$ serve as the initial state of the time evolution.
The probability amplitude for the system to be found in the state $| \phi_{n} \rangle$ at time $t  > t_0$ is given by $\langle \phi_{n} | \psi(t) \rangle$.
To transform from the Schr\"{o}dinger to the interaction picture, we use 
\begin{equation}
\langle \phi_n | \psi(t) \rangle = \langle \phi_n | e^{-i \hat{H}_0 t} \psi_{\rm I}(t) \rangle = e^{-i E_n t} \langle \phi_n | \psi_{\rm I}(t) \rangle,
\end{equation}
which implies that  the probability is preserved under this transformation: 
\begin{equation}
| \langle \phi_n | \psi(t) \rangle |^2 = | \langle \phi_n | \psi_{\rm I}(t) \rangle |^2
\end{equation}
 for all eigenstates $| \phi_n \rangle$.
Let's retain only the first order term in the perturbation series:
\begin{equation}
  \langle \phi_n | \psi_{\rm I}(t) \rangle \approx \langle \phi_n | \phi_0 \rangle -  i \lambda \int_{t_0}^{t} dt' \langle \phi_n | \hat{V}_{\rm I}(t') | \phi_0 \rangle = -i \lambda \int_{t_0}^{t} dt' \langle \phi_n | e^{i \hat{H}_{0} t'} \hat{V}(t') e^{-i \hat{H}_{0} t'} | \phi_0 \rangle.
\end{equation}
Let's assume the perturbation to be the form $\hat{V}(t) = e^{\eta t} \hat{V}$ representing a 'slow turn on', with $\eta = 0^{+}$, and $\hat{V}$ a time-independent function.
If $\eta = 0$, then the perturbation is time-independent.
If $\eta = 0^{+}$, then $e^{\eta t_0} \rightarrow 0$ as $t_0 \rightarrow - \infty$.
The construction thus effectively kills the perturbation far in the distant past, but slowly turns it on to full strength at $t = 0$.
Then, the amplitude in state $| \phi_n \rangle$ simplifies:
\begin{equation}
  \langle \phi_n | \psi_{\rm I}(t) \rangle \approx -i \lambda \int_{t_0}^{t} dt' \langle \phi_n | e^{i \hat{H}_{0} t'} e^{\eta t'} \hat{V} e^{-i \hat{H}_{0} t'} | \phi_0 \rangle = -i \lambda \langle \phi_n | \hat{V} | \phi_0 \rangle \int_{t_0}^{t} dt' e^{i(E_n - E_0)t'} e^{\eta t'},
\end{equation}
and the integral over time may be evaluated exactly to yield
\begin{equation}
  \int_{t_0}^{t} dt' e^{i(E_n - E_0)t'} e^{\eta t'} = \frac{e^{i(E_n - E_0)t} e^{\eta t} - e^{i(E_n - E_0)t_0} e^{\eta t_0}}{i (E_n - E_0) + \eta} \overset{t_0 \rightarrow - \infty}{=} \frac{e^{i(E_n - E_0)t} e^{\eta t}}{i (E_n - E_0) + \eta}.
\end{equation}
The amplitude then is 
\begin{equation}
  \langle \phi_n | \psi_{\rm I}(t) \rangle \approx -i \lambda \langle \phi_n | \hat{V} | \phi_0 \rangle \cdot \frac{e^{i(E_n - E_0)t} e^{\eta t}}{i (E_n - E_0) + \eta} = \lambda \langle \phi_n | \hat{V} | \phi_0 \rangle \cdot \frac{e^{i(E_n - E_0)t} e^{\eta t}}{(E_n - E_0) - i\eta}.
\end{equation}
The probability of the state making a transition from $| \phi_0 \rangle$ to $| \phi_n \rangle$ at time $t$ is
\begin{equation}
  | \langle \phi_n | \psi(t) \rangle |^2 = | \langle \phi_n | \psi_{\rm I}(t) \rangle |^2 \approx \lambda^2 | \langle \phi_n | \hat{V} | \phi_0 \rangle |^2 \frac{e^{2 \eta t}}{(E_0 - E_n)^2 + \eta^2}.
\end{equation}
The rate of transitions from state $| \phi_0 \rangle \rightarrow | \phi_n \rangle$ is
\begin{equation}\label{eq:rate}
  \frac{1}{\tau_{|\phi_0\rangle \rightarrow |\phi_n\rangle}} = \frac{d}{dt} | \langle \phi_n | \psi(t) \rangle |^2 \approx \lambda^2 | \langle \phi_n | \hat{V} | \phi_0 \rangle |^2 \frac{2 \eta}{(E_0 - E_n)^2 + \eta^2} e^{2 \eta t}.
\end{equation}
When $\eta \rightarrow 0$, Eq.~\eqref{eq:rate} is 0, except when $E_0 - E_n = 0$.
Taking advantage of the identity $ \lim_{\eta \rightarrow 0^{+}} \frac{1}{i}[\frac{1}{x - i \eta} - \frac{1}{x + i \eta}] = 2 \pi \delta(x)$,
the rate of transitions can be written as 
\begin{equation}
  \frac{1}{\tau_{|\phi_0\rangle \rightarrow |\phi_n\rangle}} \approx 2 \pi \lambda^2 | \langle \phi_n | \hat{V} | \phi_0 \rangle |^2 \delta(E_0 - E_n),
\end{equation}
which is the Fermi's golden rule.
The rate of the transitions from $| \gamma \rangle = \sum_{n} c_n | \phi_n \rangle$ to $| \Gamma \rangle = \sum_{n} d_n | \phi_n \rangle$ can be described as 
\begin{equation}
  \frac{1}{\tau_{| \gamma \rangle \rightarrow | \Gamma \rangle}} = \sum_{n , m} |c_n|^2 |d_m|^2 \frac{d}{dt} | \langle \phi_n | \Phi_m(t) \rangle |^2 \approx 2 \pi \lambda^2 \sum_{n , m} |c_n|^2 |d_m|^2 | \langle \phi_n | \hat{V} | \phi_m \rangle |^2 \delta(E_n - E_m),
\end{equation}
where $\Phi_m(t)$ is the time-evolved state of $| \phi_m \rangle$ in the interaction picture.
According to the Fermi's golden rule, we can obtain the thermalization time bound $t^{*} \sim \mathcal{O}(\lambda^{-2})$.

\section{Thermalization time bound of the thermal states}
Consider a Hamiltonian $\hat{H}_{0}$ perturbed by time-dependent term $\hat{V}(t)$ with small perturbation strength $\lambda$.
The full Hamiltonian is given by $\hat{H} = \hat{H}_{0} + \lambda \hat{V}$. 
$|\phi_{n}\rangle$ is an eigenstates of $\hat{H}_{0}$ with the corresponding eigenstates $E_{n}$. The eigenstate thermalization hypothesis (ETH) suggests that in the non-integrable quantum systems, if the physical expectation values $\langle \phi_n | \hat{A} | \phi_n \rangle$ of the system's Hamiltonian eigenstates $| \phi_n \rangle$ smoothly depend on energy with minimal fluctuations near the energy $E_n$:
\begin{equation}
  (\Delta E)^2 |A''(E)/A(E)| \ll 1,
\end{equation} 
the system will tend toward thermal equilibrium after long-term dynamical evolution, and its long-term state can be described by an ensemble~\cite{rigol_thermalization_2008},
where $E$ is the width of the energy distribution in the ensemble and $A(E)$ is the dependence of the expectation value of the observable $A_{nn} = \langle \phi_n | \hat{A} | \phi_n \rangle$ on the energy $E_n$.
If one takes $\hat{\rho}$ and evolves it under the unitary evolution dictated by $\hat{H}_{0}$, observables equilibrate to the predictions of the so-called diagonal ensemble $\hat{\rho}_{\textrm{DE}}$~\cite{mallayya_prethermalization_2019}:
\begin{equation}
  \hat{\rho}_{\textrm{DE}} = \sum_{n} (\langle \phi_n | \hat{\rho} | \phi_n \rangle) | \phi_n \rangle \langle \phi_n |.
\end{equation}
The expectation value of the observable $\hat{O}$ is $O_{\textrm{DE}} = \textrm{Tr}[\hat{O} \hat{\rho}_{\textrm{DE}}]$, when the system tends to thermal equilibrium.
Assume that the effect of the perturbation is small at times $t \ll 1/\lambda$, in the sense that, for such times, one can meaningfully approximate the dynamics under $\hat{H}$ by the reference dynamics generated by $\hat{H}_{0}$. During this process, we can observe a fast initial relaxation of observables toward the expectation value predicted by the diagonal ensemble~\cite{mallayya_prethermalization_2019, PhysRevB.104.184302}, which is so-called the prethermalization.
 
Based on Fermi's golden rule, we can obtain the rate at which the true equilibrium is approached:
\begin{equation}
  \frac{dO}{dt} = 2 \pi \lambda^2 \sum_{n , m} | \langle \phi_m | \hat{V} | \phi_n \rangle |^2 \delta(E_n - E_m) (\langle \phi_m | \hat{O} | \phi_m \rangle - \langle \phi_n | \hat{O} | \phi_n \rangle) \langle \phi_n | \hat{\rho}(t) | \phi_n \rangle + \mathcal{O}(\lambda^3). 
\end{equation}
The thermalization time bound of the thermal states derived is $t^{*} \sim \mathcal{O}(\lambda^{-2})$.

\end{document}